\documentclass[%
 reprint,
 amsmath,amssymb,
 aps,
pra,
]{revtex4-2}

\usepackage{graphicx}
\usepackage{dcolumn}
\usepackage{bm}

\usepackage{diagbox}
\usepackage{multirow}
\usepackage{hyperref}

\usepackage{xcolor}

\usepackage{CJKutf8}
\usepackage[T1]{fontenc}

\begin{document}

\begin{CJK}{UTF8}{gbsn}

\title{A decomposition for transverse spins in structured vector fields}
\author{{Zhi-Kang Xiong}\textsuperscript{1}}
\author{{Zhen-Lai Wang}\textsuperscript{2}}
\author{{Y. Liu (刘泱杰)}\textsuperscript{1, 3}} 
\email{Corresponding author: yangjie@hubu.edu.cn}
\author{{Meng Wen}\textsuperscript{1}}
\author{{Bin Zhou}\textsuperscript{1, 4}}
\affiliation
{
$^{1}$ Department of Physics, School of Physics, Hubei University, Wuhan 430062, China\\
$^{2}$ School of Mathematics and Physics, Hubei Polytechnic University, Huangshi 435003, China\\
$^{3}$ Lanzhou Center for Theoretical Physics, Key Laboratory of Theoretical Physics of Gansu Province, and Key Laboratory of Quantum Theory and Applications of MoE, Lanzhou University, Lanzhou 730000, China\\
$^{4}$ Wuhan Institute of Quantum Technology, Wuhan 430206, China
}

%

\begin{abstract}

Classical vector waves can possess intricate spin angular momenta (SAM), which are \emph{perpendicular} to the propagation direction, as revealed by the recent recognition of surprisingly transverse SAM in electromagnetic (EM) fields. In this paper, we employ the Hertz potential method to define structured vector fields and analytically decompose the SAM of the wave fields in two parts. Our novel approach of decomposition not only confirms that transverse SAM may originate from the first-order spatial inhomogeneity of the Poynting momentum, but also points out that for \emph{non-planar vector waves with near fields}, an extraordinary spin appears as a distinct part out of transverse spin. By four examples of vector beams, we further demonstrate that the proposed transverse spins prevail universally in both propagating and evanescent waves. This work renews our fundamental understanding of the decomposition of SAM for classical vector waves. 

\end{abstract}

\maketitle

\end{CJK}


\section{\label{sec:level1}Introduction}
Light carries angular momenta related to the rotational, spatial, and polarization distributions of the electromagnetic (EM) waves~\cite{Jackson1999, Allen2016, Andrews2012}, which respectively boils down to orbital angular momenta (OAM) with helical phases, and spin angular momenta (SAM) with circular polarizations. Of the whole SAM for EM waves, the longitudinal part is recognized as an average of spins of photons contained in the beam, i.e. $\mathbf{s}_{\rm l}\propto \hbar\langle\mathbf{k}\sigma\rangle$, with $\mathbf{k}$ the wave vector, $\hbar$ Planck constant and $\sigma=\pm 1$ representing the spin up and down respectively~\cite{Griffiths2018, Vzutic2004}. Therefore, the vector beams with left and right circular polarization are conventionally characterized by the longitudinal spins $\mathbf{s}_{\rm l}\propto \pm\mathbf{k}$ respectively~\cite{Griffiths2013, Ishimaru2017, Bliokh2015spin}. 

Other than the longitudinal SAM $\mathbf{s}_{\rm l}$ parallel to the direction of wave propagation, the transverse SAM $\mathbf{s}_{\rm t}$ was found to direct orthogonal to the propagation direction in the surface wave context~\cite{Bliokh2012, Bliokh2014, Bekshaev2015}, which has enriched the scope of SAM for classical waves. This transverse SAM was found ubiquitous in structured wave fields, such as evanescent waves, focused beams, interference fields, etc.~\cite{Aiello2015transverse, Bliokh2014, Bekshaev2015, Bliokh2015Transverse, Aiello2015}. Due to its ability to control unidirectional propagation via spin~\cite{Oconnor2014, Mitsch2014}, the transverse spin has found a wide range of application prospects, utilized in optical nano-fibers or photonic crystal waveguides~\cite{Francisco2013, Mitsch2014, Petersen2014, Deng2017}. In addition, transverse SAM of a light field rotates a trapped particle upon it, which provides additional rotational degrees of freedom for use in optical tweezers~\cite{Pal2020, ShiY2023}. Furthermore, the emergent onset of transverse spin–momentum locking~\cite{Bliokh2015Transverse, Van2016, Bliokh2015quantum, ShiP2023Dynamical}, became readily applicable in fields of quantum information~\cite{Petersen2014, Le2015} and
topological photonics~\cite{Lu2014, Gustafsson2000}. 

In 2021, Shi~\emph{et al.}~\cite{ShiP2021transverse} characterized that the transverse spin of the linearly polarized evanescent wave with the kinetic momentum $\textbf{p}$ for wave fields, as $\textbf{s}_{\rm t}\propto\nabla\times\textbf{p}$. This formulation contributes to the design of electromagnetic guided waves with specific transverse SAM, and it was further used to dissect transverse SAM in a similar manner to that of the Poynting momentum, in propagation wave
fields later on~\cite{ShiP2021transverse, ShiP2022}. We note that this notion is based on angular spectrum expansion~\cite{SongC2025Generic} in that a curl of the kinetic vector directs perpendicular to the wave vector, and the general scenario which may deviate from such due to the chaotic multiple components, is thus restrained solely for evanescent or guided waves~\cite{Bekshaev2022t-spin}. Henceforth, it remains curious to test whether such a notion of transverse SAM generally apply to generic non-planar vector waves including propagating waves even in free space. 

In this paper, we apply the Hertz vector method to build structured waves by exploiting the polarization degree of freedom, and decompose the SAMs for both \emph{evanescent} and \emph{propagating} waves. In general, the SAM of a vector wave is found to divide into two parts: one is caused by the Poynting spin $\textbf{s}_{\rm p}$ due to the spatial inhomogeneity of momentum density, and the other originates from the higher order of spatial variance for fields, which we name as $\textbf{s}_{\rm e}$ throughout our paper. As examples, we calculate the spin densities of a few structured vector waves, including cosine evanescent beam, the Bessel beam, Gaussian beam and Airy beam with elliptic polarization. To note, our decomposition of SAM differs from previous works~\cite{ShiP2021transverse, ShiP2022, Vernon2024}, and our newly-defined extraordinary spin is intimately connected to \emph{the polarization degree-of-freedom, near field, and the non-planar wave front} for vector waves.

The rest of our paper is structured as follows: in Sec.~\ref{sec:level2}, we calculate the spin decompositions for both evanescent and propagating wave fields in the general form defined by the Hertz vectors. In Sec.~\ref{level3}, we present the concrete decompositions of transverse SAMs for exemplary wave fields in sequence: cosine evanescent wave, Bessel propagating wave, Gaussian beam and Airy beam. Moreover, we discuss about possible experimental platform to test our theory. In Sec.~\ref{remarks} we remark about the experimental discussion and an extension to scalar waves for our theory meant for vector waves. And the summary is put in Sec.~\ref{conclusion}. 

\section{\label{sec:level2}Basic derivations}

\subsection{\label{Subsec-2A}Vector wave equations and the decomposition \\formulae of spins}
The Hertz vector method is a useful mathematical tool for building vector wave solutions~\cite{Essex1977, Stratton2007, Ornigotti2014, Wang2014}. Assuming a time-harmonic term $e^{-i \omega t}$, 
the wave equation for the Hertz vector can be written as:
\begin{eqnarray}
\nabla^2\bm{\Pi}+k^2\bm{\Pi}=0. \label{01}
\end{eqnarray}
Here, the Hertz vector $\bm{\Pi}=\Pi(x, y, z)\hat{n}$, where $k=\omega\sqrt{\epsilon_0\mu_0}$ is the wave number, $\epsilon_0$ and $\mu_0$ are the vacuum permittivity and permeability respectively. The direction $\hat{n}$ of the Hertz vector can be any constant unit vector. The wave equation Eq.~\eqref{01} has two independent solutions $\bm{\Pi}_e$ and $\bm{\Pi}_m$, which are known as the electric and the magnetic Hertz potentials~\cite{Jackson1999, Stratton2007} respectively. The hence-defined vector solutions are E-type and H-type waves, respectively as
\begin{eqnarray}
\mathrm{E-type}:\qquad  \textbf{E}&=&\nabla(\nabla\cdot\bm{\Pi}_e)+k^2\bm{\Pi}_e, \label{2}\\
\textbf{H}&=&-i\omega\epsilon_0\nabla\times\bm{\Pi}_e, \label{3}\\
\mathrm{H-type}:\qquad  \textbf{E}&=&i\omega\mu_0\nabla\times\bm{\Pi}_m, \label{4}\\
\textbf{H}&=&\nabla(\nabla\cdot\bm{\Pi}_m)+k^2\bm{\Pi}_m. \label{5}
\end{eqnarray}
The E-type and H-type vector waves are dually symmetric, so only one is sufficient to define the electromagnetic field~\cite{Ornigotti2014, Wang2014}. Here throughout our paper we choose $\bm{\Pi}_m$ to represent EM waves. 

For a vector wave field, the time-averaged energy density $w$ and the time-average kinetic momentum (or Poynting vector) $\textbf{p}$ can be expressed by EM fields, as 
\begin{eqnarray}
w=\frac{1}{4}(\epsilon_0\textbf{E}^*\cdot\textbf{E}+\mu_0\textbf{H}^*\cdot\textbf{H}), \label{6}
\end{eqnarray}
and 
\begin{eqnarray}
\textbf{p}=\frac{\epsilon_0\mu_0}{2}\textrm{Re}[\textbf{E}^*\times\textbf{H}]. \label{7}
\end{eqnarray}
Here, the momentum can be separated into orbital momentum $\textbf{p}_{\rm o}$ and spin momentum $\textbf{p}_{\rm s}$, i.e. $\textbf{p}=\textbf{p}_{\rm o}+\textbf{p}_{\rm s}$~
\cite{Berry2009Opt_currents, Bekshaev2013Sub, Bliokh2016, MaoY2021}, where
\begin{eqnarray}
\textbf{p}_{\rm o}&=&\frac{1}{4\omega}\textrm{Im}[\epsilon_0\textbf{E}^*\cdot(\nabla)\textbf{E}+\mu_0\textbf{H}^*\cdot(\nabla)\textbf{H}], \label{8}\\
\textbf{p}_{\rm s}&=&\frac{1}{2}\nabla\times\textbf{s}, \label{9}
\end{eqnarray}
and $\textbf{s}$ denotes the time-averaged SAM in EM fields, 
\begin{eqnarray}\label{9a}
\textbf{s}=\frac{1}{4\omega}\textrm{Im}[\epsilon_0\textbf{E}^*\times\textbf{E}+\mu_0\textbf{H}^*\times\textbf{H} ]. 
\end{eqnarray}
In EM fields, the transverse constraint $\nabla\cdot\textbf{E}=0$ results in an imaginary longitudinal electric field, which component is $\pi/2$ out of phase with respect to the transverse one~\cite{Richards1959EM-diffraction}. This phase difference implies the occurrence of a transversely spinning field in the propagation plane, which contains the direction of wave propagation~\cite{Bliokh2015Transverse, Aiello2015}. In this case, the electric field vector rotates about an axis perpendicular to the plane of propagation, resulting in spin components perpendicular to the plane of propagation. Note that our framework adopts the electric–magnetic democracy to give a symmetric form for fields ~\cite{Bekshaev2011Internal, Berry2009Opt_currents} guaranteed by electromagnetic duality symmetry~\cite{Fernandez2013EM_duality}.  


We now give the central result of our paper in general form: the decomposition of spins in analogous to that of momentum in Eqs.~\eqref{8} and ~\eqref{9}. By substituting the curl relations in Maxwell equations in Eq.~\eqref{9a}, the SAM for vector EM waves $\textbf{s}$ can be decomposed to two parts, $\textbf{s}_{\rm p}$ and $\textbf{s}_{\rm e}$ (cf. Sec.~$\mathrm{I}$ of Supplemental Material~\cite{SM} for their explicit derivation) as
\begin{eqnarray}\label{10}
\textbf{s}=\textbf{s}_{\rm p}+\textbf{s}_{\rm e},
\end{eqnarray}
with
\begin{eqnarray}
\textbf{s}_{\rm p}&=&\frac{1}{k^2}\nabla\times\textbf{p},\label{12}\\
\textbf{s}_{\rm e}&=&\frac{1}{4\omega k^2}\textrm{Im}[\epsilon_0\nabla \mathbf{E}^*\times (\cdot \nabla ) \mathbf{E}-\epsilon_0  \nabla \mathbf{E}^* \cdot (\times  \nabla ) \mathbf{E} \nonumber \\
&+&\mu_0\nabla \mathbf{H}^*\times (\cdot \nabla ) \mathbf{H} 
-\mu_0\nabla \mathbf{H}^* \cdot (\times \nabla ) \mathbf{H}],  \label{13}
\end{eqnarray}
where the double operators $\times\cdot$ and $\cdot\times$ for vectors $\mathbf{A}$ and $\mathbf{B}$ are defined respectively as
\begin{eqnarray}\label{double1}
\nabla\mathbf{A}\times(\cdot \nabla)\mathbf{B}\equiv \hat{l}\epsilon_{lmn}\partial_p A_m \partial_pB_n, 
\end{eqnarray}
and 
\begin{eqnarray}\label{double2}
\nabla\mathbf{A}\cdot (\times \nabla)\mathbf{B}\equiv \hat{l}\epsilon_{lmn}\partial_m A_p \partial_n B_p, 
\end{eqnarray}
in component form under Einstein convention, with $\epsilon_{lmn}$ as Levi-Civita symbol. Equation~\eqref{12} corresponds to the spin part caused by inhomogeneity of kinetic momentum $\textbf{s}_{\rm p}$~\cite{Vernon2024}~\footnote{Note that our $\textbf{s}_{\rm p}$ is defined twice of that  from ~\cite{Vernon2024}. }, and equation~\eqref{13} describes the extraordinary part $\textbf{s}_{\rm e}$ arising from higher order derivatives of the wave fields. Using Eq.~\eqref{10}, we decompose their two parts of spins for various wave fields to reveal the composition of transverse SAM for various types of wave beams in free space. This decomposition in Eq.~\eqref{12} is intensionally to take the unity curl of the momentum, instead of a half in Ref.~\cite{ShiP2021transverse}, which thus identifies the extraordinary spin in Eq.~\eqref{13} as a measure to the polarization degree-of-freedom, near field, and the non-planar wave front for vector waves. We remark that such spin momenta also contain topological and dynamical textures~\cite{ShiP2023Dynamical}. 

\subsection{\label{Subsec-2B}General theory of electromagnetic \\spin decomposition}

In this subsection, we give general formulae for spin decomposition of both evanescent and propagating waves respectively. Firstly, we consider an evanescent wave decaying along $z$ direction in the general sense. Secondly the propagating wave in $z$ direction is also treated. 

Firstly for evanescent wave, the Hertz vector is given by a solution to wave equation Eq.~\eqref{01} as 
\begin{eqnarray}
\bm{\Pi}_m=F(x,y)e^{-\kappa z}\hat{n}, \label{14}
\end{eqnarray}
where the complex function $F(x,y)$ satisfies $\nabla^2_{\bot} F + \beta^2 F = 0 $ with $\nabla^2_\bot={\partial^2}/{\partial x^2}+{\partial^2}/{\partial y^2}, \beta^2=k^2+\kappa^2$. Here $\kappa$ is the decaying wave number of the evanescent wave and $\hat{n}$ unit vector.

We choose the Hertz vector in Eq.~\eqref{14} to direct along $z$ direction, i.e. $\hat{n}=\hat{z}$, and substitute it into Eqs.~\eqref{4} and \eqref{5} to obtain the fields of the evanescent wave 
\begin{eqnarray}
\textbf{E}&=&i\omega\mu_0\left(\begin{array}{ccc}
\frac{\partial F}{\partial y}\\
-\frac{\partial F}{\partial x}\\
0\end{array}\right)e^{-\kappa z},
\\
\textbf{H}&=&\left(\begin{array}{ccc}
-\kappa\frac{\partial F}{\partial x}\\
-\kappa\frac{\partial F}{\partial y}\\
\beta^2 F\end{array}\right)e^{-\kappa z}. \label{16}
\end{eqnarray}
Using Eqs.~\eqref{7} and ~\eqref{9a}, we further obtain
\begin{eqnarray}
\textbf{p}&=&\frac{\mu_0k^2}{2\omega}e^{-2\kappa z}\textrm{Re}\left(\begin{array}{ccc}
i\beta^2 F\frac{\partial F^*}{\partial x}\\
i\beta^2 F\frac{\partial F^*}{\partial y}\\
i\kappa(\frac{\partial F^*}{\partial y}\frac{\partial F}{\partial y}+\frac{\partial F^*}{\partial x}\frac{\partial F}{\partial x})\end{array}\right),\label{17}
\end{eqnarray}
\begin{eqnarray}
\textbf{s}&=&\frac{\mu_0}{2\omega}e^{-2\kappa z}\textrm{Im}\left(\begin{array}{ccc}
\kappa\beta^2 F^*\frac{\partial F}{\partial y}\\
\kappa\beta^2 F\frac{\partial F^*}{\partial x}\\
\beta^2\frac{\partial F}{\partial y}\frac{\partial F^*}{\partial x}\end{array}\right). \label{17a}
\end{eqnarray}
The SAM $\mathbf{s}$ is separated into the Poynting spin and the extraordinary spin, each of which is respectively 
\begin{eqnarray}
\label{17b}
\textbf{s}_{\rm p}&=&\frac{\mu_0}{\omega}e^{-2\kappa z}\textrm{Im}\left(\begin{array}{ccc}
\kappa\beta^2 F^*\frac{\partial F}{\partial y}\\
\kappa\beta^2 F\frac{\partial F^*}{\partial x}\\
\beta^2\frac{\partial F}{\partial y}\frac{\partial F^*}{\partial x}\end{array}\right),
\\
\label{17c}
\textbf{s}_{\rm e}&=&-\frac{\mu_0}{2\omega}e^{-2\kappa z}\textrm{Im}\left(\begin{array}{ccc}
\kappa\beta^2 F^*\frac{\partial F}{\partial y}\\
\kappa\beta^2 F\frac{\partial F^*}{\partial x}\\
\beta^2\frac{\partial F}{\partial y}\frac{\partial F^*}{\partial x}\end{array}\right). 
\end{eqnarray}
From Eq.~\eqref{17b} above for an evanescent wave whose Hertz vector decays in its own direction, like Eq.~\eqref{14}, the total spin is exactly half the Poynting spin as the extraordinary spin takes twice the Poynting spin, i.e. 
\begin{equation}\label{twofold}
\textbf{s}=\frac{\textbf{s}_{\rm p}}{2}=-\textbf{s}_{\rm e}, 
\end{equation}
consistent with Ref.~\cite{ShiP2021transverse}. 

Moreover, tuning the direction of the Hertz vector leads to various vector polarization of fields. For example, by choosing $\hat{n}=\hat{x}$ in Eq.~\eqref{14}, one similarly obtains the electromagnetic vector field:  
\begin{eqnarray}
\textbf{E}=i\omega\mu_0\left(\begin{array}{ccc}
0\\
-\kappa F\\
-\frac{\partial F}{\partial y}\end{array}\right)e^{-\kappa z}, 
\\
\textbf{H}=\left(\begin{array}{ccc}
\frac{\partial^2 F}{\partial x^2}+k^2 F\\
\frac{\partial^2 F}{\partial y\partial x}\\
-\kappa\frac{\partial F}{\partial x}\end{array}\right)e^{-\kappa z}. \nonumber\\
\label{19}
\end{eqnarray}
Furthermore, its momentum and SAM are obtained respectively as 
\begin{eqnarray}
\label{20}
\textbf{p}=\frac{\mu_0}{2\omega}k^2e^{-2\kappa z}\textrm{Im}\left(\begin{array}{ccc}
\kappa^2 F^{*}\frac{\partial F}{\partial x}+\frac{\partial^2 F}{\partial y\partial x}\frac{\partial F^{*}}{\partial y}\\
-\frac{\partial F^{*}}{\partial y}(\frac{\partial^2 F}{\partial x^2}+k^2 F)\\
\kappa F^{*}(\frac{\partial^2 F}{\partial x^2}+k^2 F)
\end{array}\right), 
\end{eqnarray}
and
\begin{eqnarray}
\label{20a}
\textbf{s}=\frac{\mu_0}{2\omega}e^{-2\kappa z}\textrm{Im}\left(\begin{array}{ccc}
k^2\kappa F^{*}\frac{\partial F}{\partial y}+\kappa\frac{\partial^2 F}{\partial y\partial x}\frac{\partial F^{*}}{\partial x}\\
\kappa\frac{\partial F}{\partial x}(\frac{\partial^2 F^*}{\partial x^2}+k^2 F^*)\\
\frac{\partial^2 F}{\partial y\partial x}(\frac{\partial^2 F^*}{\partial x^2}+k^2 F^*)
\end{array}\right). 
\end{eqnarray}
And the SAM of electromagnetic fields Eq.~\eqref{19} is separated into
\begin{eqnarray}
\textbf{s}_{\rm p}&=&\frac{\mu_0}{2\omega}e^{-2\kappa z}\textrm{Im}\left(\begin{array}{cccc}
\kappa F^{*}\frac{\partial^3 F}{\partial y\partial x^2}+2k^2\kappa F^{*}\frac{\partial F}{\partial y}\\
-\kappa\frac{\partial F^*}{\partial y}\frac{\partial^2 F}{\partial x^2}\\
\\
-\kappa\frac{\partial F^{*}}{\partial x}\frac{\partial^2 F}{\partial x^2}-\kappa F^*\frac{\partial^3 F}{\partial x^3}\\
-2\kappa^3 F^*\frac{\partial F}{\partial x}-2\kappa\frac{\partial F^*}{\partial y}\frac{\partial^2 F}{\partial y\partial x}\\
\\
2\frac{\partial^2 F}{\partial y\partial x}(\frac{\partial^2 F^*}{\partial x^2}+k^2 F^*)-\beta^2\frac{\partial F^*}{\partial y}\frac{\partial F}{\partial x}\\
-\frac{\partial F^*}{\partial y}(\frac{\partial^3 F}{\partial x^3}+\frac{\partial^3 F}{\partial x\partial y^2})
\end{array}\right), \nonumber\\
\label{21a}
\end{eqnarray}
and
\begin{eqnarray}
\textbf{s}_{\rm e}&=&\frac{\mu_0}{2\omega}e^{-2\kappa z}\textrm{Im}\left(\begin{array}{ccc}
\kappa\frac{\partial^2 F}{\partial y\partial x}\frac{\partial F^{*}}{\partial x}-\kappa F^{*}\frac{\partial^3 F}{\partial y\partial x^2}\\
-k^2\kappa F^{*}\frac{\partial F}{\partial y}+\kappa\frac{\partial F^*}{\partial y}\frac{\partial^2 F}{\partial x^2}\\
\\
\kappa F^*\frac{\partial^3 F}{\partial x^3}+(\beta^2+\kappa^2)\kappa F^*\frac{\partial F}{\partial x}\nonumber\\
+2\kappa\frac{\partial F^*}{\partial y}\frac{\partial^2 F}{\partial y\partial x}\\
\\
-\frac{\partial^2 F}{\partial y\partial x}(\frac{\partial^2 F^*}{\partial x^2}+k^2 F^*)+\beta^2\frac{\partial F^*}{\partial y}\frac{\partial F}{\partial x}\\
+\frac{\partial F^*}{\partial y}(\frac{\partial^3 F}{\partial x^3}+\frac{\partial^3 F}{\partial x\partial y^2})
\end{array}\right), \nonumber\\
\label{21b}
\end{eqnarray}
which indicate that for evanescent waves Eq.~\eqref{19} with Hertz potential in $\hat{n}=\hat{x}$, the transverse spin contains the extraordinary part, distinct from the Poynting spin $\textbf{s}_{\rm p}$. Likewise for $\hat{n}=\hat{y}$, we calculate and get similar results to Eqs.~\eqref{21a} and~\eqref{21b}. Therefore, only when the direction of the Hertz vector aligns with the decaying direction for $\bm{\Pi}_m= F(x,y)e^{-\kappa z}\hat{z}$, is the total spin fully characterized by the Poynting momentum Eq.~\eqref{twofold}. We then conclude that for such a vector evanescent wave, one has to consider the contribution of its extraordinary spin $\mathbf{s}_{\rm e}$ in Eq.~\eqref{13}. 

Secondly, for a propagating wave we also examine its general spin decomposition and see whether extraordinary spin still exist when the Hertz vector takes
\begin{eqnarray}
\bm{\Pi}_m=\psi(x,y)e^{ik_zz}\hat{n}, \label{22}
\end{eqnarray}
where $\psi(x,y)$ fulfills $\nabla^2_\bot \psi+k_r^2 \psi=0$ with $k_r^2=k^2-k_z^2$, $k$ is the wave number. 

For $\hat{n}=\hat{z}$ the electromagnetic fields of the vector propagating wave are
\begin{eqnarray}
\textbf{E}=i\omega\mu_0\left(\begin{array}{ccc}
\frac{\partial \psi}{\partial y}\\
-\frac{\partial \psi}{\partial x}\\
0\end{array}\right)e^{ik_zz}, \label{23}
\\
\textbf{H}=\left(\begin{array}{ccc}
ik_z\frac{\partial \psi}{\partial x}\\
ik_z\frac{\partial \psi}{\partial y}\\
k_r^2\psi\end{array}\right)e^{ik_zz}.\label{24}
\end{eqnarray}
One obtains the SAM by inserting Eqs.~\eqref{23} and \eqref{24} into Eq.~\eqref{9a},
\begin{eqnarray}
\textbf{s}&=&\frac{\mu_0}{2\omega}\textrm{Re}\left(\begin{array}{cccc}
-k_zk_r^2\psi\frac{\partial\psi^*}{\partial y}\\
k_zk_r^2\psi\frac{\partial\psi^*}{\partial x}\\
-i(k^2+k_z^2)\frac{\partial\psi^*}{\partial x}\frac{\partial\psi}{\partial y}
\end{array}\right). \label{25}
\end{eqnarray}
From Eqs.~\eqref{7}, ~\eqref{12} and~\eqref{13}, one has
\begin{eqnarray}
\textbf{p}&=&\frac{\mu_0}{2\omega}k^2\textrm{Re}\left(\begin{array}{cccc}
ik_r^2\psi\frac{\partial\psi^*}{\partial x}\\
ik_r^2\psi\frac{\partial\psi^*}{\partial y}\\
k_z(\frac{\partial\psi}{\partial x}\frac{\partial\psi^*}{\partial x}+\frac{\partial\psi^*}{\partial y}\frac{\partial\psi}{\partial y})
\end{array}\right), \label{26}
\end{eqnarray}
and also 
\begin{eqnarray}
\textbf{s}_{\rm p}&=&\frac{\mu_0}{\omega}\textrm{Re}\left(\begin{array}{cccc}
k_z(\frac{\partial\psi}{\partial x}\frac{\partial^2\psi^*}{\partial x\partial y}+\frac{\partial\psi}{\partial y}\frac{\partial^2\psi^*}{\partial y^2})\\
-k_z(\frac{\partial\psi}{\partial x}\frac{\partial^2\psi^*}{\partial x^2}+\frac{\partial\psi}{\partial y}\frac{\partial^2\psi^*}{\partial y\partial x})\\
-ik_r^2\frac{\partial\psi^*}{\partial x}\frac{\partial\psi}{\partial y}
\end{array}\right), \label{25a}
\end{eqnarray}
and
\begin{eqnarray}
\textbf{s}_{\rm e}&=&\frac{\mu_0}{2\omega}\textrm{Re}\left(\begin{array}{cccc}
k_z(-k_r^2\psi\frac{\partial\psi^*}{\partial y}-2\frac{\partial\psi}{\partial x}\frac{\partial^2\psi^*}{\partial x\partial y}-2\frac{\partial\psi}{\partial y}\frac{\partial^2\psi^*}{\partial y^2})\\
k_z(k_r^2\psi\frac{\partial\psi^*}{\partial x}+2\frac{\partial\psi}{\partial x}\frac{\partial^2\psi^*}{\partial x^2}+2\frac{\partial\psi}{\partial y}\frac{\partial^2\psi^*}{\partial y\partial x})\\
-i(2k^2_z-k^2_r)\frac{\partial\psi^*}{\partial x}\frac{\partial\psi}{\partial y}
\end{array}\right).\nonumber\\
\label{25b}
\end{eqnarray}

Moreover, for the propagating wave with $\hat{n}=\hat{x}$, one gets
\begin{eqnarray}
\label{25g}
\textbf{E}=-\omega\mu_0\left(\begin{array}{ccc}
0\\
k_z \psi \\
-i\frac{\partial \psi}{\partial y}\end{array}\right)e^{ik_zz}, 
\\
\textbf{H}=\left(\begin{array}{ccc}
\frac{\partial^2 \psi}{\partial x^2}+k^2\psi\\
\frac{\partial^2\psi}{\partial y \partial x}\\
ik_z\frac{\partial \psi}{\partial x}\end{array}\right)e^{ik_zz}. \label{25c}
\end{eqnarray}
Then one obtains the momentum of the propagating wave
\begin{eqnarray}\label{25d}
\textbf{p}&=&\frac{\mu_0}{2\omega}k^2\textrm{Im}\left(\begin{array}{cccc}
k^2_z\psi^*\frac{\partial\psi}{\partial x}+\frac{\partial\psi^*}{\partial y}\frac{\partial^2\psi}{\partial y\partial x}\\
-\frac{\partial\psi^*}{\partial y}(\frac{\partial^2\psi}{\partial x^2}+k^2\psi)\\
ik_z\psi^*(\frac{\partial^2\psi}{\partial x^2}+k^2\psi)
\end{array}\right), 
\end{eqnarray}
and the SAM
\begin{eqnarray}
\label{SAMprop}
\textbf{s}&=&\frac{\mu_0}{2\omega}\textrm{Im}\left(\begin{array}{cccc}
ik_z(k^2\psi^*\frac{\partial\psi}{\partial y}+\frac{\partial\psi}{\partial x}\frac{\partial^2\psi^*}{\partial x\partial y})\\
-ik_z\frac{\partial\psi^*}{\partial x}(\frac{\partial^2\psi}{\partial x^2}+k^2\psi)\\
\frac{\partial^2\psi}{\partial y\partial x}(\frac{\partial^2\psi^*}{\partial x^2}+k^2\psi^*)
\end{array}\right), 
\end{eqnarray}
which can be divided into the Poynting spin and the extraordinary spin respectively, 
\begin{eqnarray}\label{25f}
\textbf{s}_{\rm p}&=&\frac{\mu_0}{2\omega}\textrm{Im}\left(\begin{array}{cccc}
ik_z(\frac{\partial\psi^*}{\partial y}\frac{\partial^2\psi}{\partial x^2}+2k^2\psi\frac{\partial\psi^*}{\partial y}+\psi^*\frac{\partial^3\psi}{\partial y\partial x^2})\\
-ik_z(\frac{\partial\psi^*}{\partial x}\frac{\partial^2\psi}{\partial x^2}+2k^2\psi\frac{\partial\psi^*}{\partial x}+\psi^*\frac{\partial^3\psi}{\partial x^3})\\
2\frac{\partial^2\psi}{\partial y^2}\frac{\partial^2\psi^*}{\partial x\partial y}-2k^2_z\frac{\partial\psi^*}{\partial y}\frac{\partial\psi}{\partial x}
\end{array}\right), \nonumber
\\
\end{eqnarray}
and
\begin{eqnarray}\label{25e}
\textbf{s}_{\rm e}&=&\frac{\mu_0}{2\omega}\textrm{Im}\left(\begin{array}{cccc}
ik_z(\frac{\partial\psi}{\partial x}\frac{\partial^2\psi^*}{\partial x\partial y}-\frac{\partial\psi^*}{\partial y}\frac{\partial^2\psi}{\partial x^2}-k^2\psi\frac{\partial\psi^*}{\partial y}-\nonumber\\
\psi^*\frac{\partial^3\psi}{\partial y\partial x^2})\\
\\
ik_z(k^2\psi\frac{\partial\psi^*}{\partial x}+\psi^*\frac{\partial^3\psi}{\partial x^3})\\
\\
\frac{\partial^2\psi}{\partial y\partial x}(\frac{\partial^2\psi^*}{\partial x^2}+k^2\psi^*)-2\frac{\partial^2\psi}{\partial y^2}\frac{\partial^2\psi^*}{\partial x\partial y}+\nonumber\\
2k^2_z\frac{\partial\psi^*}{\partial y}\frac{\partial\psi}{\partial x}
\end{array}\right). \nonumber\\
\end{eqnarray}
 
We are then equipped with general formulae derived in this section to decompose the transverse SAM into Poynting spin and extraordinary spin jointly, for four examples of vector beams in Sec.~\ref{level3}. 

\section{Calculation and discussion}\label{level3}
On the basis of the general derivation above, in this section we give explicit SAM components for examples of structured light vector fields: cosine evanescent wave, Bessel propagating wave, Gaussian beam, and Airy beam, all of which carry non-uniform SAMs. As we will find out in each example, it is the polarization degree-of-freedom, near field, and the non-planar wave front, that distinct the extraordinary spins out of their transverse SAMs. 

\subsection{Elliptic polarization: cosine evanescent wave}\label{Cosine}
In this subsection we use a cosine evanescent wave with elliptic polarization by Hertz vector method to unveil its spin decomposition. For a cosine evanescent wave \cite{Jackson1999, ShiP2021transverse} propagating along $y$ direction and decaying in $z$ direction, its wave vector is $\textbf{k}=k_x\hat{x}+k_y\hat{y} + i\kappa\hat{z}$. We assume the Hertz potential to take this form as
\begin{eqnarray}
\bm{\Pi}_m=A_0\cos(k_xx)e^{ik_yy-\kappa z}\hat{n}, \label{27}
\end{eqnarray}
which is one solution of Hertz vector Helmholtz equation Eq.~\eqref{01} with $A_0$ as its amplitude. In the case of $\hat{n}=\hat{z}$, we calculate the fields for the cosine evanescent vector beam with elliptic polarization,
\begin{eqnarray}
\textbf{E}=A_0\omega\mu_0e^{ik_yy-\kappa z}\left(\begin{array}{ccc}
-k_y\cos(k_xx)\\
ik_x\sin(k_xx)\\
0\end{array}\right), \label{28}
\end{eqnarray}
and
\begin{eqnarray}
\textbf{H}=A_0e^{ik_yy-\kappa z}\left(\begin{array}{ccc}
k_x\kappa\sin(k_xx)\\
-ik_y\kappa\cos(k_xx)\\
\beta^2\cos(k_xx)\end{array}\right), \label{29}
\end{eqnarray}
where $\beta^2=k_x^2 + k_y^2$. Taking Eqs.~\eqref{28} and~\eqref{29} into Eqs.~\eqref{7} and ~\eqref{9a}, one gets
\begin{eqnarray}
\textbf{p}&=&\frac{\mu_0A^2_0}{2\omega}k^2e^{-2\kappa z}\left(\begin{array}{cccc}
0\\
\beta^2k_y\cos^2(k_xx)\\
0
\end{array}\right), \label{31}
\end{eqnarray}
and
\begin{eqnarray}
\textbf{s}&=&\frac{\mu_0A^2_0}{2\omega}e^{-2\kappa z}\left(\begin{array}{cccc}
\beta^2k_y\kappa\cos^2(k_xx)\\
0\\
-\beta^2k_xk_y\sin(k_xx)\cos(k_xx)
\end{array}\right). \nonumber\\
\label{30}
\end{eqnarray}
The SAM is exactly perpendicular to the direction of wave propagation, which can be attributed from a quarter $\pi/2$ phase difference between the longitudinal field component ($E_y, H_y$) and the transverse components ($E_x, H_x, H_z$), which will cause the electromagnetic field to rotate in the plane containing the wave propagation direction. From Eq.~\eqref{17c} with $F=\cos(k_xx)e^{ik_yy}$, one gets
\begin{eqnarray}
\textbf{s}_{\rm e}&=&-\frac{\mu_0A^2_0}{2\omega}e^{-2\kappa z}\left(\begin{array}{cccc}
\beta^2k_y\kappa\cos^2(k_xx)\\
0\\
-\beta^2k_xk_y\sin(k_xx)\cos(k_xx)
\end{array}\right)\nonumber\\
&=&-\frac{1}{2}\textbf{s}_p. \label{32}
\end{eqnarray}
Thus the transverse spin $\textbf{s}$ can be characterized by Poynting spin $\textbf{s}_{\rm p}$ in this case, i.e. $\textbf{s}=\textbf{s}_{\rm p}/2$, as in \cite{ShiP2021transverse}. 

When we choose the Hertz vector in Eq.~\eqref{27} to direct in $x$ direction, from Eq.~\eqref{19} its electromagnetic fields are
\begin{eqnarray}
\textbf{E}&=&A_0\omega\mu_0e^{ik_yy-\kappa z}\left(\begin{array}{ccc}
0\\
-i\kappa\cos(k_xx)\\
k_y\cos(k_xx)\end{array}\right),\label{32a}\\
\textbf{H}&=&A_0e^{ik_yy-\kappa z}\left(\begin{array}{ccc}
(k_y^2-\kappa^2)\cos(k_xx)\\
-ik_xk_y \sin(k_xx)\\
k_x\kappa\sin(k_xx)\end{array}\right),\label{32b}
\end{eqnarray}
We note that the cosine wave fields with a different Hertz direction $\hat{n}$ behave distinctly as their energy densities and momenta show in Figs.~\ref{figs-fig1}(a) and~\ref{figs-fig1}(b) (cf. Sec.~$\mathrm{II}$ of Supplemental Material~\cite{SM}), and thus they differ in both energy density and momentum.

\begin{figure}[hbtp!]
  \begin{center}
  \includegraphics[width=0.48\textwidth]{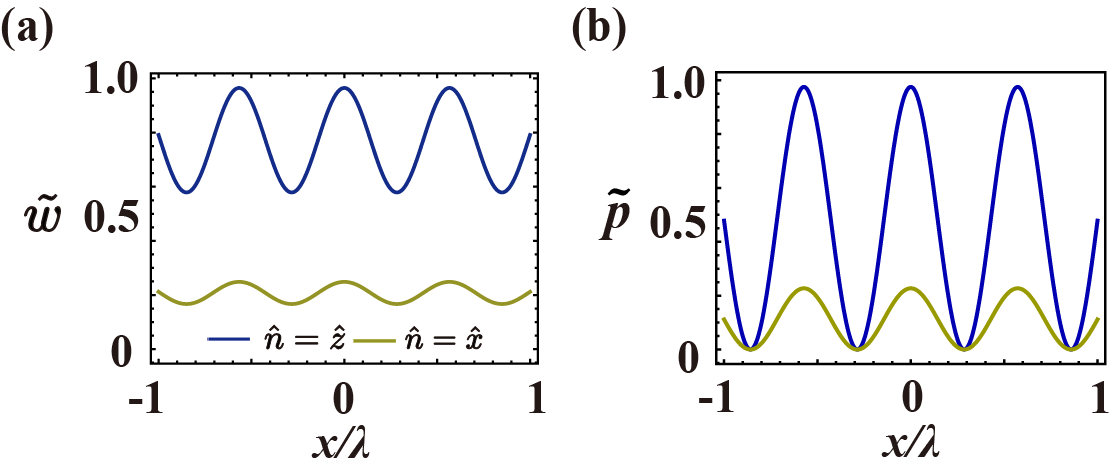}
  \caption{\label{figs-fig1}(a) The normalized energy density $\tilde{w}$ and (b) normalized momentum $\tilde{p}$ distributions of the two different fields $\hat{n}=\hat{z}$ and $\hat{n}=\hat{x}$ when $z=0$. Parameters: $k_x=\beta\sin(\pi/3)$, $k_y=\beta\cos(\pi/3)$, $\kappa=\beta\sin(\pi/16)$, $\beta=k/\cos(\pi/16)$, and $k=2\pi/\lambda$. }
  \end{center}
\end{figure}

We calculate the momentum of this field by substituting Eqs.~\eqref{32a} and ~\eqref{32b} into Eqs.~\eqref{20} and~\eqref{20a}, as 
\begin{eqnarray}
\textbf{p}&=&\frac{\mu_0A^2_0}{2\omega}k^2e^{-2\kappa z}\left(\begin{array}{cccc}
0\\
(k^2_y-\kappa^2)k_y\cos^2(k_xx)\\
0
\end{array}\right), \label{32d}
\\
\textbf{s}&=&\frac{\mu_0A^2_0}{2\omega}e^{-2\kappa z}\left(\begin{array}{cccc}
k_y\kappa [ k^2\cos^2(k_xx)+k_x^2\sin^2(k_xx) ]\\
0\\
-(k^2_y-\kappa^2)k_xk_y\sin(k_xx)\cos(k_xx)
\end{array}\right). \nonumber \\
\label{33}
\end{eqnarray}

\begin{figure*}[hbtp!]
  \includegraphics[width=0.98\textwidth]{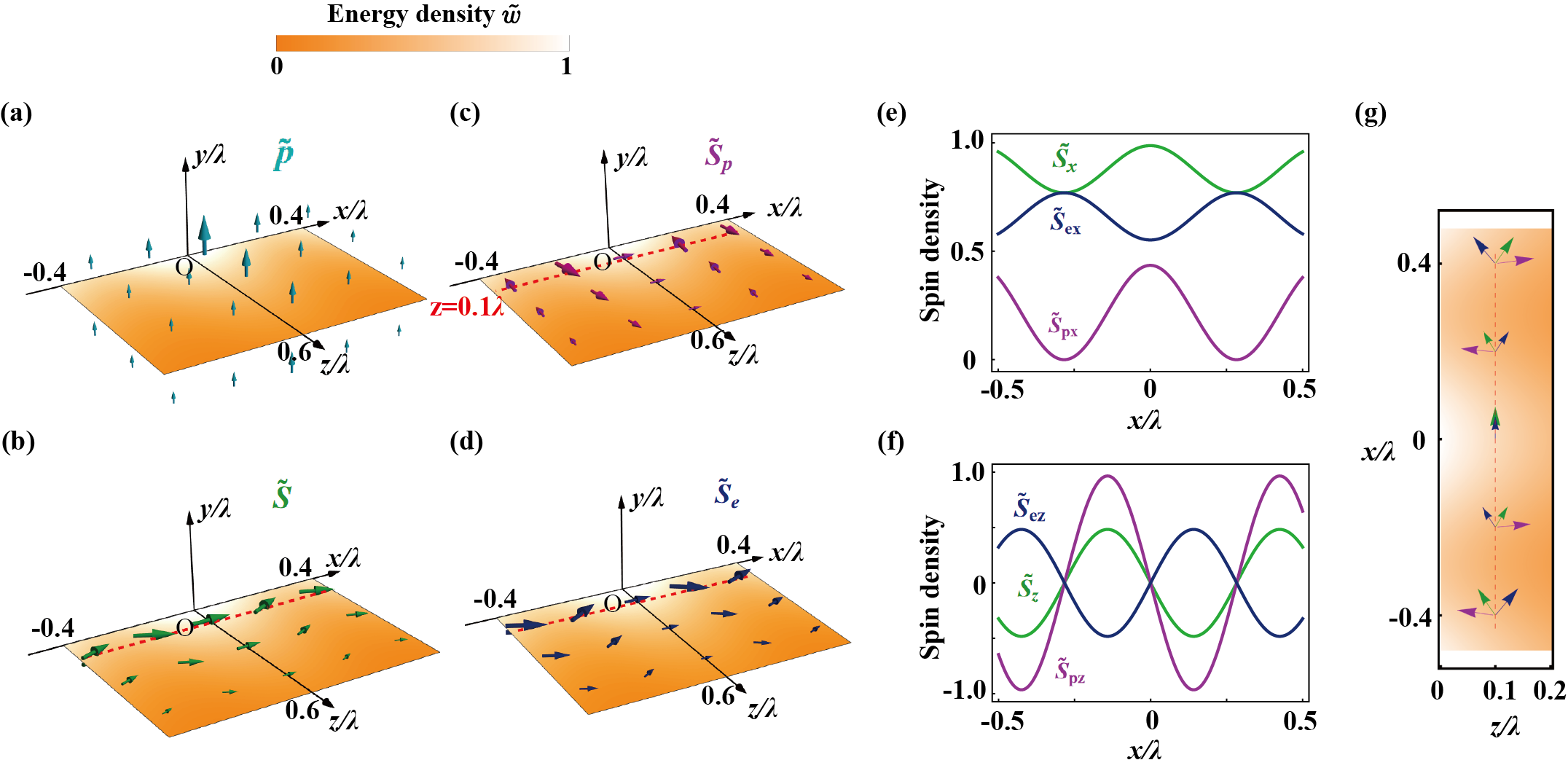}
  \caption{\label{figs-fig2}The cosine evanescent wave field defined by $\bm{\Pi}_m=\Pi_m\hat{x}$. (a) Normalized momentum $\tilde{p}$ and (b) Normalized SAM $\tilde{S}$ vector distribution. (c)-(d) Normalized Poynting spin $\tilde{S}_{\rm{p}}$ and extraordinary spin $\tilde{S}_{\rm{e}}$ vector distributions respectively. The orange background image represents the normalized energy density $\tilde{w}$ distribution in (a)-(d) and this applies for Figs. ~\ref{figs-fig3}, \ref{figs-fig6}, \ref{figs-fig8} and \ref{figs-fig9}. The red dashed line in (b)-(d) represents $z=0.1\lambda$ and $y=0$. (e)-(f) shows the curves of $x$ component and $z$ component densities of the three spins versus the lateral position $x$ when $z = 0.1\lambda$ respectively. (g) The vector distribution of three spins on plane $z=0.1\lambda$. Parameters: $k_x=\beta\sin(\pi/3)$, $k_y=\beta\cos(\pi/3)$, $\kappa=\beta\sin(\pi/16)$, $\beta=k/\cos(\pi/16)$, $k=2\pi/\lambda$.} 
\end{figure*}

The vector distributions of the normalized momentum $\tilde{p}$ and SAM $\tilde{S}$ are shown in Figs.~\ref{figs-fig2} (a) and~\ref{figs-fig2} (b) respectively. From Fig.~\ref{figs-fig2} (a) and Eq.~\eqref{32d}, the momentum vector of the wave points in $y$ direction, and its amplitude decays in $z$ direction. And from Fig.~\ref{figs-fig2} (b) and Eq.~\eqref{33}, the SAM is always pointing along positive $x$ direction while carrying a wiggling $z$ component around zero.

Taking $F=\cos(k_xx)e^{ik_yy}$ into Eqs.~\eqref{21a} and~\eqref{21b}, one obtains the decomposition of SAM for cosine evanescent beam, 
\begin{eqnarray}
  \textbf{s}_{\rm p}&=&\frac{\mu_0A^2_0}{\omega}e^{-2\kappa z}\left(\begin{array}{cccc}
    (k^2_y-\kappa^2)k_y\kappa\cos^2(k_xx)\\
    0\\
      -(k^2_y-\kappa^2)k_xk_y\sin(k_xx)\cos(k_xx)
\end{array}\right), \nonumber\\
\label{35}\\
\textbf{s}_{\rm e}&=&\frac{\mu_0A^2_0}{2\omega}e^{-2\kappa z}\left(\begin{array}{cccc}
k_y\kappa[k_x^2+(k^2_y-\kappa^2)\cos^2(k_xx)]\\
0\\
(k^2_y-\kappa^2)k_xk_y\sin(k_xx)\cos(k_xx)
\end{array}\right). \nonumber\\
\label{36}
\end{eqnarray}
The normalized two parts $\tilde{S}_p$ and $\tilde{S}_e$ vector distributions are shown in Figs.~\ref{figs-fig2}(c) and ~\ref{figs-fig2} (d), both transverse. In all four panels of Fig.~\ref{figs-fig2}, the normalized energy density $\tilde w$ of the wave field are put as background color. The red dashed line in Figs.~\ref{figs-fig2} (b)-\ref{figs-fig2}(d) represents $z=0.1\lambda$, and figures~\ref{figs-fig2} (e) and ~\ref{figs-fig2} (f) show that $x$ and $z$ components of the three spins as a function of the lateral position $x$ for $z=0.1\lambda$. Thus $x$ component of three spins in the modulation direction is non-negative while the $z$ component of three spins in the decaying direction is \emph{not}, which wiggles in the $\pm z$ direction. 

More importantly, the known double relation Eq.~\eqref{32} breaks down here for $\hat{n}=\hat{x}$ because the extraordinary spin overweighs in the modulating direction, $\mathbf{s}_{{\rm e}, x}$ in this case [cf. Fig.~\ref{figs-fig2} (e)], though the decomposition of transverse spins in the decaying direction, $\mathbf{s}_z$, still obeys the double relation [cf. Fig.~\ref{figs-fig2} (f)].  When put in vector form, the two parts weight comparable seen from the three spins at $z=0.1\lambda$ in Fig.~\ref{figs-fig2} (g). The results for the case $\hat{n}=\hat{y}$ is similar to Eqs.~\eqref{35} and~\eqref{36} (cf. Sec.~$\mathrm{II}$ of Supplemental Material~\cite{SM}). 

So for cosine evanescent wave, extraordinary spin $\mathbf{s}_{\rm e}$ generally plays a critical role in its transverse spin except when the Hertz vector points in the decaying direction, which we attribute from the elliptic polarization of vector beams out of the Hertz vector chosen. 

In addition, for propagating cosine wave, we get similar results (cf. Sec.~$\mathrm{II}$ of Supplemental Material~\cite{SM}). For another instance, we treat the elliptically polarized plane wave and a two-wave interference, whose SAM can still be divided into two parts $\textbf{s}_{\rm p}$ and $\textbf{s}_{\rm e}$ (cf. Sec.~$\mathrm{III}$ and $\mathrm{IV}$ of Supplemental Material~\cite{SM}), and breaks the previous double relation Eq.~\eqref{32}.

\subsection{Near field: Bessel propagating beam}\label{Bessel}
As a next instance we exemplify the classical diffraction-less beams in free space: the zero-order Bessel beam \cite{Wang2014, Mcgloin2005, Mishra1991, Bouchal1995}, and reveal that near fields can also contribute to the extraordinary spin. We consider the zero-order Bessel beam to intentionally exclude the spin-orbit coupling of high-order Bessel beams for they carry orbital angular momenta~\cite{ZhangX2024Ph_spin-orbit, DuL2019Deep-sub} which may complicate the scenario. The Hertz vector of the Bessel beam in cylindrical coordinates is first chosen as
\begin{eqnarray}
\bm{\Pi}_m&=&A_0J_0(rk_r)e^{ik_zz}\hat{n}, \label{37}
\end{eqnarray}
where $A_0$ is the amplitude and $J_0(rk_r)$ is the zero-order Bessel function. The wave number $k$ fulfills $k^2=k_r^2+k_z^2$. When $\hat{n}=\hat{z}$, from Eq.~\eqref{37} the electromagnetic fields are
\begin{eqnarray}
\textbf{E}=A_0\omega\mu_0e^{ik_zz}\left(\begin{array}{ccc}
-iyk_rJ_1(rk_r)/r\\
ixk_rJ_1(rk_r)/r\\
0\end{array}\right),\label{38}
\\
\textbf{H}=A_0k_re^{ik_zz}\left(\begin{array}{ccc}
-ixk_zJ_1(rk_r)/r\\
-iyk_zJ_1(rk_r)/r\\
k_rJ_0(rk_r)\end{array}\right). \label{39}
\end{eqnarray}
The electric field in Eq.~\eqref{38} carries linear polarization, and the magnetic field in Eq.~\eqref{39} carries elliptic polarization enabled by the Hertz vector method, which resemble Bessel fields in~\cite{Aiello2015transverse}. 

\begin{figure*}[htbp!]
  \includegraphics[width=0.94\textwidth]{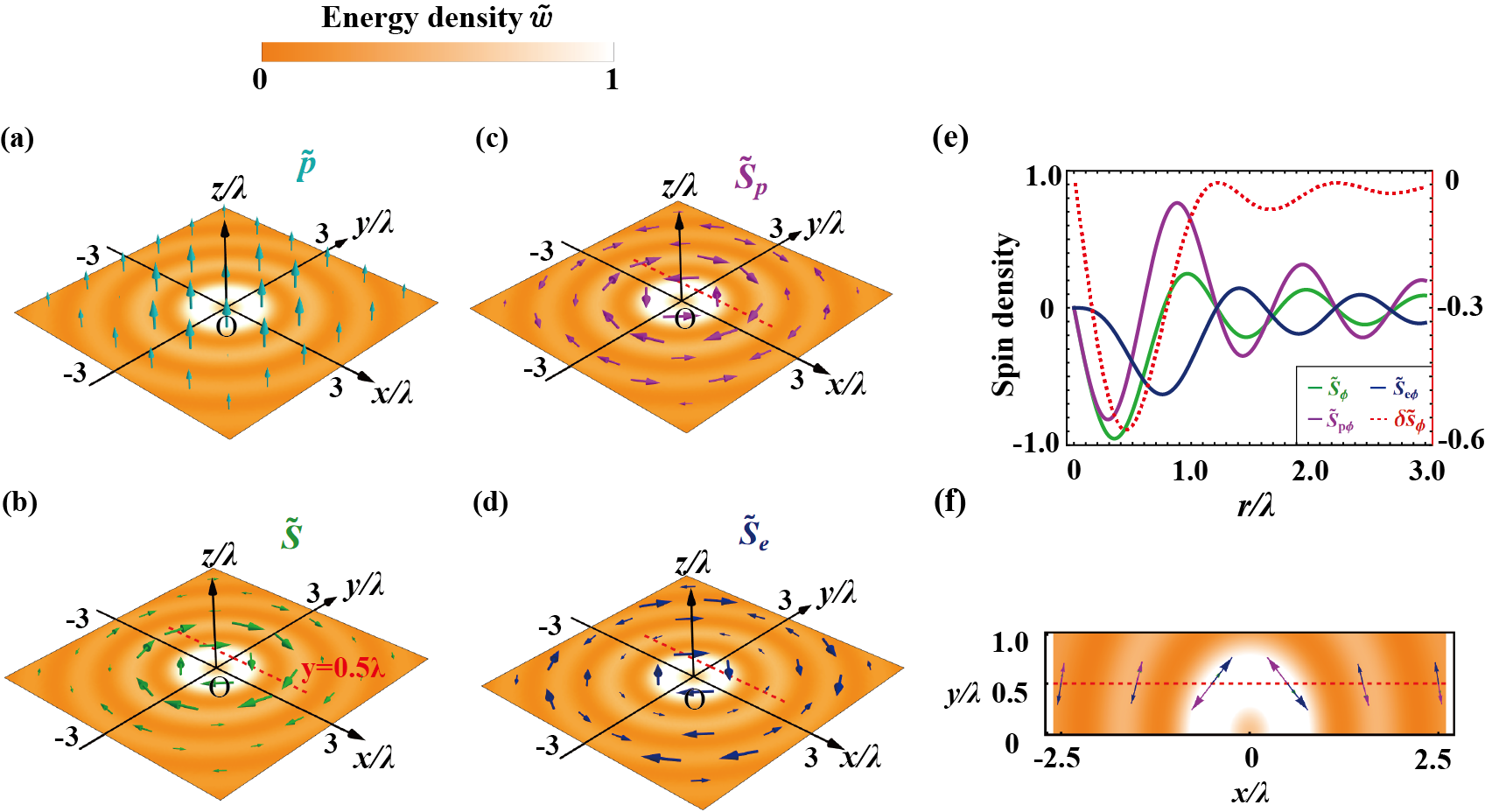}
  \caption{\label{figs-fig3}The Bessel beam defined by $\Pi_m=\Pi_m\hat{z}$. (a) Normalized momentum $\tilde{p}$ and (b) Normalized SAM $\tilde{S}$ vector distribution; (c)-(d) Normalized Poynting spin $\tilde{S}_{\rm{p}}$ and extraordinary $\tilde{S}_{\rm{e}}$ vector distributions. (e) the density distribution of three spins in $r$ direction and $\delta \tilde{s}_\phi=\tilde{S}_{\rm{p}}/2+\tilde{S}_{\rm{e}}$; (f) the spin vectors on the red dash line ($y=0.5\lambda$, $z=0$) in $x\text{-}y$ plane. Parameters: $k_r=k\cos(\pi/3)$, $k_z=k\sin(\pi/3)$, $k=2\pi/\lambda$. }
\end{figure*}

Using Eq.~\eqref{26}, one obtains the momentum of the Bessel beam
\begin{eqnarray}
\textbf{p}&=&\frac{\mu_0A^2_0}{2\omega}k^2
k^2_rk_zJ^2_1(rk_r)\hat{z}. \label{40}
\end{eqnarray}
For such a Bessel propagating wave, the momentum is purely longitudinal as expected, as Fig.~\ref{figs-fig3}(a) shows superimposed on the normalized energy density $\tilde{w}$. Also the electric field in Eq.~\eqref{38} is completely transverse, while the magnetic in Eq.~\eqref{39} is not due to the $\pi/2$ phase advance of the longitudinal component over the transverse. Thus the transverse spin results from the magnetic field: 
\begin{eqnarray}
\mathbf{s}=\textbf{s}_H = -\frac{\mu_0A^2_0}{2\omega}k^3_rk_zJ_0(rk_r)J_1(rk_r)\hat{\phi}, \label{41}
\end{eqnarray}
as shown in Fig.~\ref{figs-fig3}(b).

Similarly from Eqs.~\eqref{25a} and \eqref{25b}, the SAM for Bessel vector beam can be divided into two parts as
\begin{eqnarray}
\textbf{s}_{\rm p}&=&\frac{\mu_0A^2_0}{\omega}k^2_rk_z 
\Big[
-k_rJ_0(rk_r)J_1(rk_r)+\nonumber \\
&& \frac{J^2_1(rk_r)}{r}\Big]\hat{\phi}, \label{42}
\end{eqnarray}
and
\begin{eqnarray}\label{43}
\textbf{s}_{\rm e}&=&\frac{\mu_0A^2_0}{2\omega}k^2_rk_z
\Big[ 
k_rJ_0(rk_r)J_1(rk_r)-\nonumber\\
&& \frac{2J^2_1(rk_r)}{r} \Big] \hat{\phi}. 
\end{eqnarray}
From Eqs.~\eqref{42} and \eqref{43} Bessel beam induces an opposite near-field term in the azimuthal direction $\hat\phi$, respectively in $\mathbf{s}_{\rm p}$ [Fig.~\ref{figs-fig3} (c)] and $\mathbf{s}_{\rm e}$ [Fig.~\ref{figs-fig3} (d)], and they cancel each other when summed up in $\mathbf{s}$ [cf.~Eq.~\eqref{41}]. This near field term of squared the Bessel function, $J^2_1(rk_r)/r$ contributes to the extraordinary spin $\textbf{s}_{\rm e}$. Interestingly, the transverse spin for Bessel vector fields is dominated by the Poynting spin $\textbf{s}_{\rm p}$, as shown in radial direction in Fig.~\ref{figs-fig3} (e), and on a line $y=0.5\lambda$ in Fig.~\ref{figs-fig3} (f). The near field becomes pronounced at $r=0.6\lambda$ as $\delta \tilde{s}_\phi=\tilde{S}_{\rm{p}}/2+\tilde{S}_{\rm{e}}$ in Fig.~\ref{figs-fig3}(e) shows. And when only the far field is concerned, the squared Bessel function $J_1^2(rk_r)/r$ becomes negligible in Eqs.~\eqref{42} and \eqref{43}. Then the spins reduce to satisfy the double relationship Eq.~\eqref{twofold}.

Moreover, the Hertz vector can polarize in $x$ direction, 
\begin{eqnarray}
\bm{\Pi}_m&=&A_0 J_0(rk_r)e^{ik_zz}\hat{x}, \label{b1}
\end{eqnarray}
when its electromagnetic fields are
\begin{eqnarray}
\textbf{E}=A_0\omega\mu_0e^{ik_zz}\left(\begin{array}{ccc}
0\\
-k_zJ_0(rk_r)\\
iyk_rJ_1(rk_r)/r\end{array}\right),\label{b2}
\end{eqnarray}
and
\begin{eqnarray}
\mathbf{H}=A_0e^{ik_zz}\begin{pmatrix}
k^2J_0(rk_r)-{k_r^2x^2J_0(rk_r)}/{r^2}+\\
{k_r(x^2-y^2)J_1(rk_r)}/{r^3}\\
\\
- {k^2_rxyJ_0(rk_r)}/{r^2} + {2k_rxyJ_1(rk_r)}/{r^3} \\
\\
-{ik_rk_zxJ_1(rk_r)}/{r}
\end{pmatrix}, \nonumber\\
\label{b3}
\end{eqnarray}
with elliptic polarization. Now both the electric and magnetic fields from the Hertz vector in $x$ direction provide SAM, different from the previous case of $\bm{\Pi}_m=\Pi_m\hat{z}$. We find that 
\begin{eqnarray}
\textbf{s}_E&=&s_{E_x}\hat{x}, \label{sE}\\
\textbf{s}_H&=&s_{H_x}\hat{x}+s_{H_y}\hat{y}, \label{sH}
\end{eqnarray}
where
\begin{eqnarray}
s_{E_x}&=&-\frac{\mu_0A^2_0}{2\omega}k^2k_rk_z\frac{y}{r}J_0(rk_r)J_1(rk_r), \label{sEx}\\
\label{sHx}
s_{H_x}&=&\frac{\mu_0A^2_0}{2\omega}\Big[
k^3_rk_z\frac{x^2y}{r^3}J_0(rk_r)J_1(rk_r)-\nonumber \\
& & 2k^2_rk_z\frac{x^2y}{r^4}J^2_1(rk_r) \Big], \\
\label{sHy}
s_{H_y}&=&\frac{\mu_0A^2_0}{2\omega}\Big[
k^2 k_rk_z\frac{x}{r}J_0(rk_r)J_1(rk_r)- \nonumber \\
& & k^3_rk_z\frac{x^3}{r^3}J_0(rk_r)J_1(rk_r) + \nonumber \\
& & k^2_rk_z\frac{x(x^2-y^2)}{r^4}J^2_1(rk_r) \Big]. 
\end{eqnarray}
And they add up to its transverse spin
\begin{eqnarray}
\textbf{s}&=&\textbf{s}_E+\textbf{s}_H\nonumber\\
&\approx&\frac{\mu_0A^2_0}{2\omega}\left(\begin{array}{ccc}
-k^2k_rk_z\frac{y}{r}J_0(rk_r)J_1(rk_r)\\
k^2k_rk_z\frac{x}{r}J_0(rk_r)J_1(rk_r)\\
0\end{array}\right). \label{sEH}
\end{eqnarray}

\begin{figure}[hbtp!]
\includegraphics[width=0.48\textwidth]{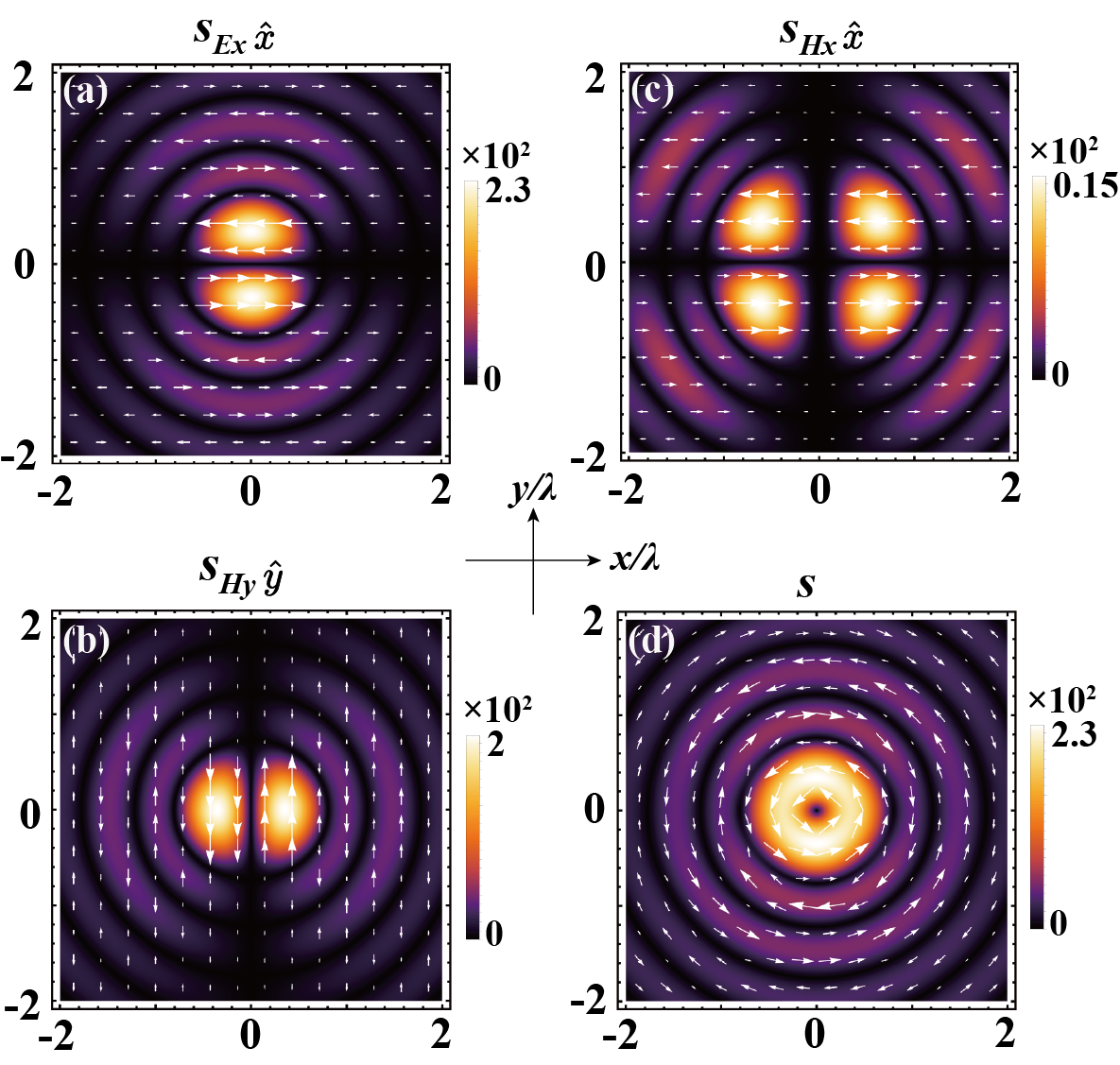}
\caption{\label{figs-fig4}The Bessel beam defined by $\bm{\Pi}=\Pi\hat{x}$. (a) Vector density distribution of SAM $\textbf{s}_E=s_{E_x}\hat{x}$ provided by the electric field; (b)-(c) distribution for two components of magnetic field $\textbf{s}_H=s_{H_x}\hat{x}+s_{H_y}\hat{y}$; (d)  distribution of the total spin $\textbf{s}=\textbf{s}_E+\textbf{s}_H$. The arrow represents the size and direction of the spin vector, the background represents the spin density. Parameters: $k_r=k\cos(\pi/3)$, $k_z=k\sin(\pi/3)$, $k=2\pi/\lambda$. } 
\end{figure}

The last equation of Eq.~\eqref{sEH} stands in far field so that $|\textbf{s}_{H_x}|\ll|\textbf{s}_{H_y}|\approx|\textbf{s}_{E_x}|$, as the higher-order terms than $1/r^2$ becomes ignorable [all the following approximation in Subsec.~\ref{Bessel} are taken the same as the far-field limit]. The spin densities of $\mathbf{s}_E$, $s_{H_x}\hat{x}$, $s_{H_y}\hat{y}$ and $\mathbf{s}$ in Eqs.~\eqref{sEx}-~\eqref{sEH} are presented in Figs.~\ref{figs-fig4} (a)- \ref{figs-fig4} (d) respectively, superimposed on a density plot of the norm of the corresponding fields. And $\textbf{s}_E$ and $\textbf{s}_H$ are nearly perpendicular to each other [cf. Figs.~\ref{figs-fig4} (a) and ~\ref{figs-fig4} (b)] and are comparable in size as expected.

The momentum of the Bessel vector beam can be obtained by taking $\psi= A_0 J_0(rk_r)$ into Eq.~\eqref{25d}, i.e.
\begin{eqnarray}
\textbf{p}&=&\frac{\mu_0A^2_0}{2\omega}k^2
\Big[k_rk_z\frac{x^2-y^2}{r^3}J_0(rk_r)J_1(rk_r)- \nonumber\\
&\quad&k^2_rk_z\frac{x^2}{r^2}J^2_0(rk_r)
+k^2k_zJ^2_0(rk_r)\Big]\hat{z}\nonumber\\
&\approx&\frac{\mu_0A^2_0}{2\omega}k^4k_zJ^2_0(rk_r)\hat{z}. \label{b7}
\end{eqnarray}
And from Eqs. \eqref{SAMprop}-~\eqref{25e}, the Poynting spin follows the double relation Eq.~\eqref{twofold} in the far-field approximation (cf. Sec.~$\mathrm{V}$ of Supplemental Material~\cite{SM}): 
\begin{eqnarray}
\textbf{s}_{\rm p}&\approx&\frac{\mu_0A^2_0}{\omega}\left(
\begin{array}{ccc}
-k^2k_rk_z\frac{y}{r}J_0(rk_r)J_1(rk_r)\\
k^2k_rk_z\frac{x}{r}J_0(rk_r)J_1(rk_r)\\
0
\end{array}\right)\nonumber\\
&\approx&2\textbf{s}. \label{b8}
\end{eqnarray}
Though the explicit expressions for the Poynting spin and the extraordinary spin are not given here to save space, all the hidden higher-order terms for them represent just the near fields. And the same results apply for the Hertz vector $\bm{\Pi}_m=\Pi_m\hat{y}$.  


So for the Bessel propagating vector beam, its transverse spin of the wave fields is decomposed of Poynting spin $\textbf{s}_{\rm p}$ and extraordinary spin $\textbf{s}_{\rm e}$, which respectively indicates the far field and the near field terms. And we infer that the near field terms are intimately connected to the extraordinary spin for this Bessel vector beam. This also applies for a higher-order Bessel beam (cf. Sec.~$\mathrm{V}$ of Supplemental Material~\cite{SM}). 

\subsection{Non-planar phase: Gaussian beam}\label{Gaussian}
Previous two examples of vector beams in Subsecs.~\ref{Cosine} and ~\ref{Bessel} propagate with planar wave fronts. Now in this subsection, we examine a typical beam carrying a non-planar wave front: the Gaussian beam under the paraxial approximation \cite{Bliokh2015Transverse, Agrawal1979, Alda2003}, which carries both evanescent and propagating waves respectively present in the previous two beams. 

Assuming the ansatz solution in a constant direction to the wave equation Eq.~\eqref{01}
\begin{eqnarray}
\bm{\Pi}=u(x,y,z)e^{ikz}\hat{n}, 
\end{eqnarray}
one obtains
\begin{eqnarray}
  \frac{\partial^2u}{\partial x^2}+\frac{\partial^2u}{\partial y^2}+\frac{\partial^2u}{\partial z^2}+2ik\frac{\partial u}{\partial z}=0. \label{44a}
\end{eqnarray}
Under the paraxial approximation $|\partial^2_zu|\ll |2k\partial_zu|$ and $|\partial^2_zu|\ll|\partial^2_xu|,|\partial^2_yu|$, a solution of Gaussian beam can be written as
\begin{eqnarray}
u(x,y,z)=A_0\frac{z_R}{z-iz_R}\exp\frac{ikr^2}{2(z-iz_R)}, \label{57}
\end{eqnarray}
where $A_0$ is a constant field amplitude, $z_R={kw^2_0}/{2}$ is the Rayleigh diffraction length, $w_0=w(z=0)$ is the beam waist for $w(z)=w_0\sqrt{1+(z/z_R)^2}$~\cite{Alda2003} with $r^2=x^2+y^2$. 
The Hertz vector of such a field is
\begin{eqnarray}
\bm{\Pi}_m=A_0\frac{z_R}{z-iz_R}e^{\frac{ikr^2}{2(z-iz_R)}}e^{ikz}\hat{n}.
\label{58}
\end{eqnarray}

\begin{figure}[htbp!]
  \includegraphics[width=0.48\textwidth]{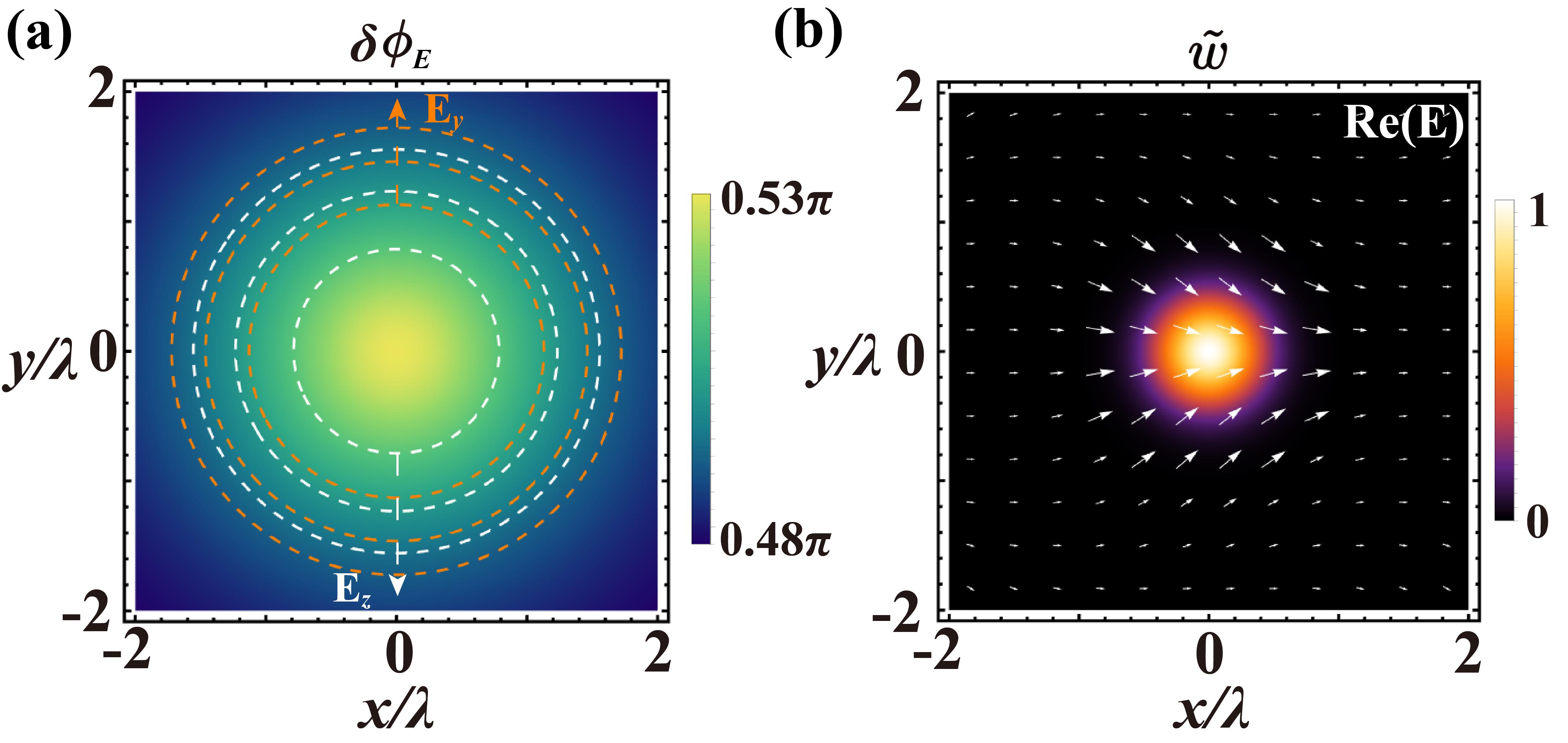}
  \caption{\label{figs-fig5}(a) The phase difference $\delta_{\phi_E}=\phi_{E_y}-\phi_{Ez}$ distribution between the two components of the Gaussian electric field defined by Eq.~\eqref{58}. From Eq.~\eqref{59a}, $E_z=0$ for plane $y=0$, so we remove the factor $y$ here. The orange and white dashed line represent the wave front of $E_y$ and $E_z$ respectively, which can also characterize the phase difference between the two components; (b) The electric field vector $\text{Re} (\mathbf{E} )$ and the energy density $\tilde{w}$ of such a Gaussian beam distribution. The white arrow represents the electric vector and the background represents the normalized energy distribution. Here, we choose the position $z=0.1\lambda$ near focus, with $z_R=\lambda$ and $k=2\pi/\lambda$.}
\end{figure}

When the Hertz vector directs in $\hat{n}=\hat{x}$, one gets the electromagnetic fields under the paraxial approximation by taking Eq.~\eqref{58} into Eqs.~\eqref{25g} and~\eqref{25c}, 
\begin{eqnarray}
\textbf{E}&=&\omega\mu_0A_0e^{i \alpha}
\left(\begin{array}{ccc}
0\\
\frac{-kz_R}{z-iz_R}-\frac{iz_R}{(z-iz_R)^2}+\frac{kr^2z_R}{2(z-iz_R)^3}\\
\frac{kyz_R}{(z-iz_R)^2}\end{array}\right),\nonumber\\ \label{59a}
\\
\mathbf{H} &=& A_0 e^{i \alpha}
\left( \begin{array}{ccc}
\frac{k^2z_R}{z-iz_R} + \frac{ikz_R}{(z-iz_R)^2}-\frac{k^2x^2z_R}{(z-iz_R)^3}\\
-\frac{k^2 xy z_R}{(z-iz_R)^3} \\
-\frac{k^2 x z_R}{(z-iz_R)^2}-\frac{2ikx z_R}{(z-iz_R)^3} +\frac{k^2 r^2x z_R}{2(z-iz_R)^4}\end{array}
\right).\nonumber\\ \label{60a}
\end{eqnarray}
where $\alpha=kz + {kr^2}/{2(z - iz_R)} $. And the phase difference of the electric field near the focal plane ($z=0.1\lambda$), spreads around $\pi/2$ unequally for two components $E_y, E_z$ as expected to make the transverse SAM, shown in Fig.~\ref{figs-fig5} (a). Substituting Eqs.~\eqref{59a} and \eqref{60a}  into Eq.~\eqref{6}, one obtains the energy density $w$. Here, we use the symbol $\approx$ to indicate that we ignore all terms of ${z^{-5}}$ and higher orders under the paraxial approximation in following calculations.

\begin{eqnarray}
w\approx\frac{\mu_0A^2_0 k^4}{2} \frac{z^2_R}{z^2+z^2_R}e^{\zeta}, \label{59}
\end{eqnarray}
where $\zeta=-kr^2z_R/(z^2+z^2_R)$. The energy density is shown in Fig.~\ref{figs-fig5} (b) [normalized $\tilde{w}$]. Using the electromagnetic fields Eqs.~\eqref{59a} and \eqref{60a}, the momentum $\textbf{p}$ and SAM $\textbf{s}$ of the Gaussian beam are [cf. Sec.~$\mathrm{VI}$ of Supplemental Material~\cite{SM} for detailed calculation results.]

\begin{eqnarray}
\textbf{p}&\approx&\frac{\mu_0A^2_0}{2\omega}k^2e^{\zeta}
\Biggl\{
\frac{k^3xzz^2_R}{(z^2+z^2_R)^2}\hat{x}+
\frac{k^3yzz^2_R}{(z^2+z^2_R)^2}\hat{y}+\nonumber\\
&&\Big[\frac{k^3z^2_R}{z^2+z^2_R}+\frac{kz^2_R}{(z^2+z^2_R)^2}-\frac{2k^2z^3_R}{(z^2+z^2_R)^2}-\nonumber\\
&&\frac{k^3r^2z^2z^2_R}{2(z^2+z^2_R)^3}-\frac{k^3x^2z^2z^2_R}{(z^2+z^2_R)^3}\Big]\hat{z}
\Biggr\}, 
\end{eqnarray}
and
\begin{eqnarray}
\textbf{s}&\approx&\frac{\mu_0A^2_0}{2\omega}k^2e^{\zeta}
\Biggl[
(\frac{kyz^2_R}{(z^2+z^2_R)^2}-\frac{k^2yz^3_R}{(z^2+z^2_R)^2})\hat{x}+\nonumber\\
&&(\frac{k^2xz^3_R}{(z^2+z^2_R)^2}-\frac{kxz^2_R}{(z^2+z^2_R)^2}+\frac{2k^3xz^2z^2_R}{(z^2+z^2_R)^3})\hat{y}
\Biggr], \nonumber\\
\label{61}
\end{eqnarray}

In order to illustrate the spin feature for this Gaussian vector field, we discuss how they behave as the position $z$ moves in and off the focal plane. 

\begin{figure*}[hbtp!]
  \includegraphics[width=0.98\textwidth]{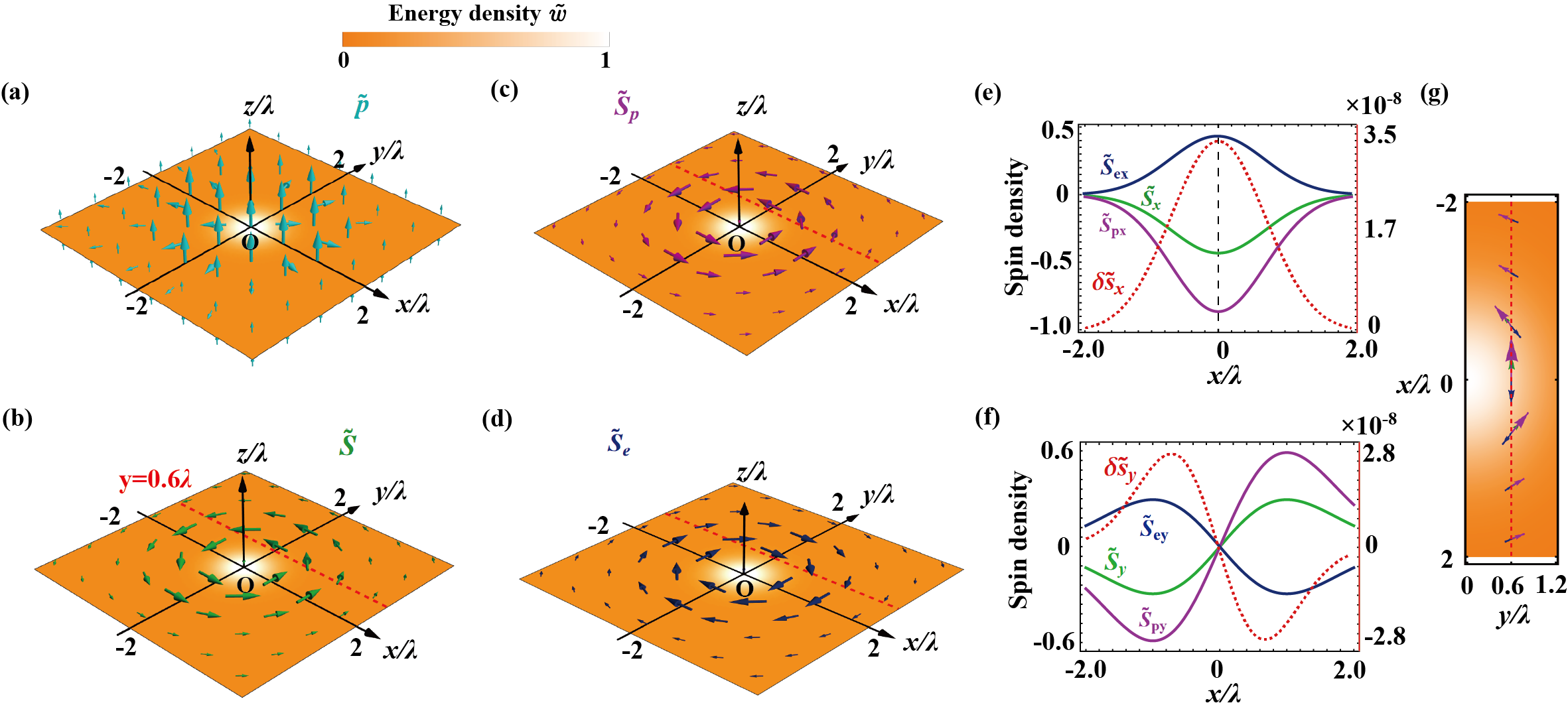}
  \caption{\label{figs-fig6}
The Gaussian beam defined by $\Pi_m=\Pi_m\hat{x}$ when $z=z_R$. (a)-(b) Normalized momentum $\tilde{p}$ total (transverse) spin $\tilde{S}$ vector distributions; (c)-(d) Normalized Poynting spin $\tilde{S}_{\rm p}$ and $\tilde{S}_{\rm e}$ vector distributions. (e)-(f) The magnitude of the $x$ and $y$ components of three spins when $y=0.6\lambda, z=z_R$ with $\delta \tilde{s}=\tilde{S}_{\rm{p}}/2+\tilde{S}_{\rm{e}}$, on the line $y=0.6\lambda, z=z_R$. The black dashed line represents $x=0$. (g) The vector distribution of three spins on the line $y=0.6\lambda, z=z_R$. Parameters: $z=z_R=\pi\lambda$ and $k=2\pi/\lambda$. }
\end{figure*}

\emph{Near the focal plane ($z\approx 0$) and its vicinity}. From Eq.~\eqref{61}, only the longitudinal momentum $p_z$ remains unvanished and the transverse spin in azimuthal direction \cite{Bliokh2015Transverse}, 
\begin{eqnarray}
\textbf{p}&\approx&\frac{\mu_0A^2_0}{2\omega}\frac{k^5z^2_R}{z^2+z^2_R}e^{\zeta}
\hat{z}, \label{61a}\\
\textbf{s}&\approx&\frac{\mu_0A^2_0}{2\omega}\frac{k^4z^2_R}{z^2+z^2_R}e^{\zeta}
\frac{rz_R}{z^2+z^2_R}
\hat{\phi}.\label{61b}
\end{eqnarray}

According to Eqs.~\eqref{61a} and \eqref{61b}, one gets the Poynting spin $\textbf{s}_{\rm p}$ and the extraordinary spin $\textbf{s}_{\rm e}$, as
\begin{eqnarray}
\textbf{s}_{\rm p}&=&\frac{\mu_0A^2_0}{2\omega}\frac{k^4z^2_R}{z^2+z^2_R}e^{\zeta}
\frac{2rz_R}{z^2+z^2_R}\hat{\phi},\label{62}\\
\textbf{s}_{\rm e}&=&-\frac{\mu_0A^2_0}{2\omega}\frac{k^4z^2_R}{z^2+z^2_R}e^{\zeta}
\frac{rz_R}{z^2+z^2_R}\hat{\phi}. \label{63}
\end{eqnarray}
So near the focal plane where the phase front stays planar, transverse spin obeys the double relation Eq.~\eqref{twofold}.

\emph{Off the focal plane $z\gg 0$}, i.e. \emph{$z\gtrsim z_R$}. Then the contributions of the transverse components $p_r$ of the wave field momentum becomes critical
\begin{eqnarray}
\textbf{p}&\approx&\frac{\mu_0A^2_0}{2\omega}\frac{k^5z^2_R}{z^2+z^2_R}e^{\zeta}\big(
\frac{r z}{z^2+z^2_R}\hat{r}+\hat{z}
\big)
\label{64}
\end{eqnarray}
and its SAM becomes
\begin{eqnarray}\label{65}
\textbf{s}&\approx&\frac{\mu_0A^2_0}{2\omega}\frac{k^4z^2_R}{z^2+z^2_R}e^{\zeta}
\frac{rz_R}{z^2+z^2_R}\hat{\phi}, 
\end{eqnarray} 
both of which are shown in Figs.~\ref{figs-fig6} (a) and \ref{figs-fig6} (b). Due to the radial momentum $p_r$, the Poynting spin $\textbf{s}_{\rm p}$ and $\textbf{s}_{\rm e}$ are distinct, 
\begin{eqnarray}
\textbf{s}_{\rm p}&=&\frac{\mu_0A^2_0}{2\omega}\frac{k^4z^2_R}{z^2+z^2_R}e^{\zeta}\Biggl[\frac{2r z_R}{z^2+z^2_R}+\frac{r}{k(z^2+z^2_R)}\Biggr]\hat{\phi},\nonumber\\
\\
\textbf{s}_{\rm e}&=&\frac{\mu_0A^2_0}{2\omega}\frac{k^4z^2_R}{z^2+z^2_R}e^{\zeta}\Biggl[-\frac{r z_R}{z^2+z^2_R}-\frac{r}{k(z^2+z^2_R)}\Biggr]\hat{\phi}.\nonumber\\
\label{67}   
\end{eqnarray}
shown in Figs.~\ref{figs-fig6} (c) and \ref{figs-fig6} (d). The transverse spins $\textbf{s}$ in Eqs.~\eqref{65} and ~\eqref{67} breaks the double relationship Eq.~\eqref{twofold}, as $\delta \tilde{s}=\tilde{S}_{\rm{p}}/2+\tilde{S}_{\rm{e}}$ shows in the dashed red curves of Figs.~\ref{figs-fig6} (e)-~\ref{figs-fig6} (g). Thus we infer that the extraordinary spin results from the curved front off the focal plane. This scenario also works in the Gaussian field defined by the Hertz vector $\bm{\Pi}_m=\Pi_m\hat{y}$.

An interesting result occurs when the Hertz vector is chosen in the longitudinal direction $\bm{\Pi}_m=\Pi_m\hat{z}$. Then its electromagnetic fields are
\begin{eqnarray}
\textbf{E}=\omega\mu_0A_0e^{i\alpha}\left(\begin{array}{ccc}
-\frac{kyz_R}{(z-iz_R)^2}\\
\frac{kxz_R}{(z-iz_R)^2}\\
0\end{array}\right), \label{68}
\end{eqnarray}
and
\begin{eqnarray}
\textbf{H}=-A_0e^{i\alpha}\left(\begin{array}{ccc}
\frac{k^2xz_R}{(z-iz_R)^2}+\frac{2ikxz_R}{(z-iz_R)^3}-\frac{k^2r^2xz_R}{2(z-iz_R)^4}\\
\frac{k^2yz_R}{(z-iz_R)^2}+\frac{2ikyz_R}{(z-iz_R)^3}-\frac{k^2r^2yz_R}{2(z-iz_R)^4}\\
\frac{2ikz_R}{(z-iz_R)^2}-\frac{k^2r^2z_R}{(z-iz_R)^3}\end{array}\right).\nonumber\\ \label{69}
\end{eqnarray}
The magnetic field $\textbf{H}$ satisfies the wave function under paraxial approximation.

Using Eqs.~\eqref{68} and~\eqref{69}, one gets the energy $w$, momentum $\textbf{p}$ and SAM $\textbf{s}$ of the field respectively [cf. Sec.~$\mathrm{VI}$ of Supplemental Material~\cite{SM}], 
\begin{eqnarray}
w&\approx&\frac{\mu_0A^2_0}{2}k^4 \frac{r^2z^2_R}{(z^2+z^2_R)^2} e^ {\zeta},\\
\textbf{p}&\approx& \frac{w}{\omega} k\hat{z}, \\
\textbf{s}&\approx& -\frac{w}{\omega} \frac{2}{kr}\hat{\phi}, \label{71}
\end{eqnarray}
which gives its momentum in the longitudinal direction [cf. Sec.~$\mathrm{VI}$ of Supplemental Material~\cite{SM}]. The azimuthal SAM $\textbf{s}$ is provided solely by the magnetic field, i.e. $\textbf{s}=\textbf{s}_H$. And the transverse spin is found to be only the Poynting spin $\textbf{s}_{\rm p}$ under paraxial approximation, 
\begin{eqnarray}
\textbf{s} \approx  \textbf{s}_{\rm p}, \label{72}
\end{eqnarray}
which is unique in that the SAM is entirely provided by the Poynting spin and no extraordinary spin exists $\textbf{s}_{\rm e}=0$ regardless the longitudinal position $z$ near or off the focus. This special result for Gaussian beam with its Hertz vector in $z$, supports our decomposition as a pure Poynting SAM instance, which can be chosen in experimental verification (also see Subec.~\ref{experiment}). 

From the Gaussian vector beam, the extraordinary spin $\textbf{s}_{\rm e}$ results from the non-planar wave front, i.e. when $z$ takes a position far away from the focus.

\subsection{Polarization, near field, and non-planar phase: Airy beam} \label{AiryBeam}
In this subsection, we consider a vector form of Airy beam, whose trajectory will bend constantly in space to give a non-planar wave front similar to the Gaussian beam in Subsec.~\ref{Gaussian}. The Airy vector beam carries both propagating and evanescent waves, which is similar to cosine evanescent wave and Bessel beam in Subsecs.~\ref{Cosine} and \ref{Bessel}. Thus the Airy vector beam carries all the three features present in the previous three examples. 

Under the paraxial approximation $|\partial^2_zu|\ll |2k\partial_zu|$, and $|\partial^2_zu|\ll|\partial^2_xu|,|\partial^2_yu|$, equation~\eqref{44a} can be written as
\begin{eqnarray}
i\frac{\partial u}{\partial\xi}+\frac{1}{2}\frac{\partial^2u}{\partial s^2}=0, \label{44}
\end{eqnarray}
where $u(s, \xi)$ is the wave envelope, $s=x/x_0$ represents a dimensionless transverse coordinate, $x_0$ is a transverse scale, and $\xi = z/kx^2_0$ is a scale-free propagation distance, $k$ is the wave number~\cite{Siviloglou2007, Siviloglou2007observation}. The Airy beam is one solution to Eq.~\eqref{44}, 
\begin{eqnarray}
u(x, z)=A_0\textrm{Ai}\Big(s-\frac{\xi^2}{4}\Big)e^{\frac{is\xi}{2}-\frac{i\xi^3}{12}}, \label{45}
\end{eqnarray}
where $\rm{Ai}(\cdot)$ is Airy function. 

Firstly, assuming $\hat{n}=\hat{y}$ the Hertz vector is
\begin{eqnarray}
\bm{\Pi}_m=A_0\textrm{Ai}\Big(s-\frac{\xi^2}{4}\Big)e^{\frac{is\xi}{2}-\frac{i\xi^3}{12}}e^{ikz}\hat{y}.\label{46}
\end{eqnarray}

\begin{figure}[hbtp!]
  \includegraphics[width=0.48\textwidth]{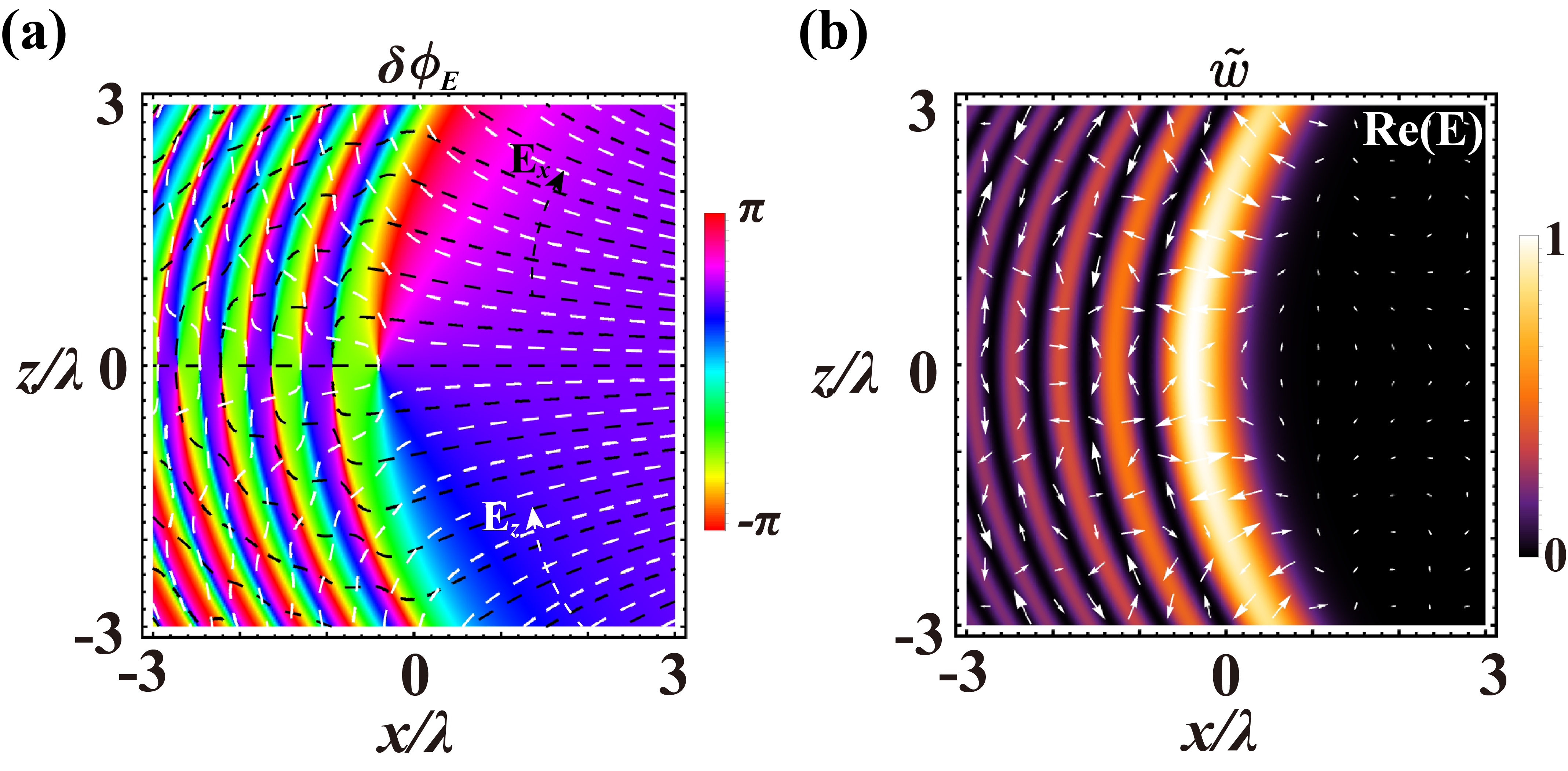}
  \caption{\label{figs-fig7} (a) The phase difference $\delta{\phi_E}=\phi_{E_x}-\phi_{Ez}$ distribution for Airy field, between the two components of the electric field defined by Eq.~\eqref{47}. The black and white dashed line represent the wave front of $E_x$ and $E_z$ respectively; (b) real part of the electric field vector $\rm Re(\textbf{E})$ superimposed on the normalized energy density $\tilde{w}$ of the Airy field. Parameters: $x_0=0.4\lambda$ and $k=2\pi/\lambda$.}
\end{figure}

Substituting Eq.~\eqref{46} into Eqs.~\eqref{4} and~\eqref{5}, we get the electromagnetic fields
\begin{eqnarray}
\textbf{E}&=&A_0\omega\mu_0e^{i\phi}\left(\begin{array}{ccc}
k\mathcal{A}+\frac{x\mathcal{A}}{2kx^3_0}-\frac{z^2\mathcal{A}}{4k^3x^6_0}+\frac{iz\mathcal{A}'}{2k^2x^4_0}\\
0\\
\frac{i\mathcal{A}'}{x_0}-\frac{z\mathcal{A}}{2kx^3_0}\end{array}\right), \nonumber \\
\label{47}
\\
\textbf{H}&=&A_0e^{i\phi}\left(\begin{array}{ccc}
0\\
k^2\mathcal{A}\\
0\end{array}\right), \label{48}
\end{eqnarray}
where $\phi={s\xi}/{2}-{\xi^3}/{12}+kz$, $\mathcal{A}=\textrm{Ai}({x}/{x_0}-{z^2}/{4k^2x^4_0})$ and $\mathcal{A}^{'}=\textrm{Ai}'({x}/{x_0}-{z^2}/{4k^2x^4_0})$ for the derivative of Airy function. We note that Eqs.~\eqref{47} and ~\eqref{48} are exact solutions to the paraxial wave equation Eq.~\eqref{44} for H-type Eqs.~\eqref{4} and~\eqref{5}. And the electric field vibrates on the propagation plane $z=0$ with a phase difference $\delta \phi_E$ fading away from the main trajectory [see Fig.~\ref{figs-fig7}] while the magnetic field linearly polarizes only in $y$ direction. Thus only the electric field contributes to the spin decomposition $\mathbf{s}_E \propto \rm{Im}\left \{ \mathbf{E}^*\times\mathbf{E}\right\}$.

The momentum $\textbf{p}$ and SAM $\textbf{s}$ of such fields can be obtained by taking Eqs.~\eqref{47} and~\eqref{48} into Eqs.\eqref{7} and~\eqref{9a}, as
\begin{eqnarray}
\textbf{p}&=&\frac{\mu_0A^2_0}{2\omega}\left(\begin{array}{ccc}
\frac{k^3z\mathcal{A}^2}{2x^3_0}\\
0\\
k^5\mathcal{A}^2+\frac{k^3x\mathcal{A}^2}{2x^3_0}-\frac{kz^2\mathcal{A}^2}{4x^6_0}\end{array}\right),\label{49}
\end{eqnarray}
and
\begin{eqnarray}
\textbf{s}&=&-\frac{\mu_0A^2_0}{2\omega}\left(\begin{array}{ccc}
0\\
\frac{k^3\mathcal{A}\mathcal{A}'}{x_0}+\frac{kx\mathcal{A}\mathcal{A}'}{2x^4_0}\\
0\end{array}\right).\label{50}
\end{eqnarray}

\begin{figure*}[hbtp!]
  \includegraphics[width=0.94\textwidth]{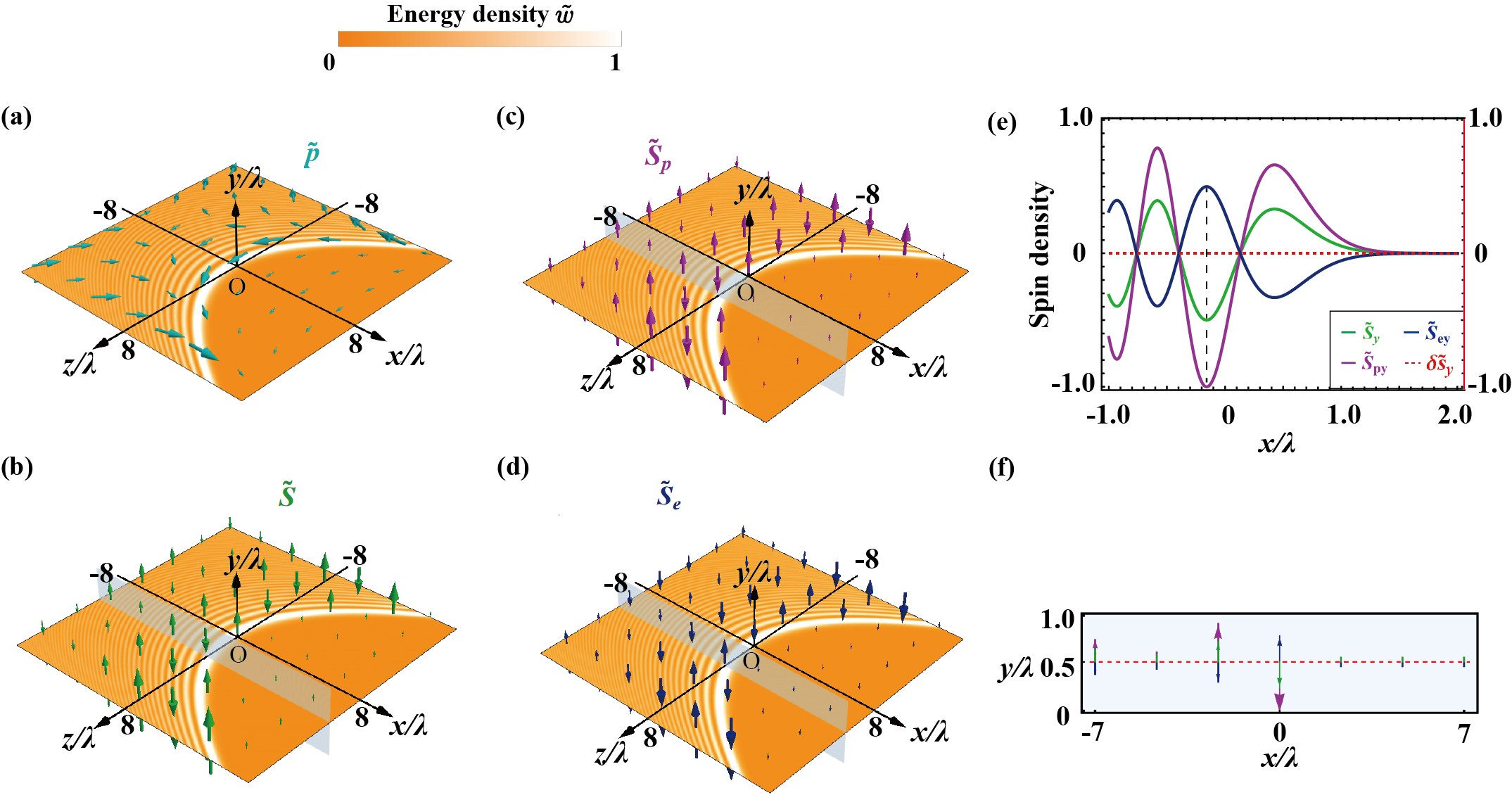}
  \caption{\label{figs-fig8} The Airy beam defined by $\bm{\Pi}_m=\Pi_m\hat{y}$. (a) Normalized momentum $\tilde{p}$ and (b) SAM $\tilde{S}$ vector distribution; (c)-(d) Normalized Poynting spin $\tilde{S}_{\rm{p}}$ and the extraordinary spin $\tilde{S}_{\rm{e}}$ vector distribution of the Airy beam. The light blue insets in (b)-(d) represent the plane at $z=2.33\lambda$. (e) The three spins density distributions, the black dashed line represents $x=-0.16\lambda$. Here $\delta \tilde{s}_y=\tilde{S}_{\rm{p}y}/2+\tilde{S}_{\rm{e}y}=0$. (f) the vector distribution of three spins at $z=2.33\lambda$ plane. Parameters: $x_0=0.4\lambda$, $k=2\pi/\lambda$.}
\end{figure*}

The momentum and transverse spin above indicate a beam propagating on $x\text{-}z$ plane: moving along the $z$ direction defined by Eq.~\eqref{46} and bending transversely towards $\pm x$ direction during propagation, as is typically known for Airy beam. This bending feature is also reflected in Fig.~\ref{figs-fig8} (a) and ~\ref{figs-fig8} (b).

Further, the transverse spin is divided into two parts $\textbf{s}_{\rm p}$ and $\textbf{s}_{\rm e}$, shown in Figs.~\ref{figs-fig8}(c) and \ref{figs-fig8} (d) respectively and the double relation Eq.~\eqref{twofold} holds when Hertz vector points along the perpendicular $y$ direction:

\begin{eqnarray}
\textbf{s}_{\rm e}&=&\frac{\mu_0A^2_0}{2\omega}\left(\begin{array}{ccc}
0\\
\frac{k^3\mathcal{A}\mathcal{A}'}{x_0}+\frac{kx\mathcal{A}\mathcal{A}'}{2x^4_0}\\
0
\end{array}\right)\nonumber\\
&=&-\frac{1}{2}\textbf{s}_{\rm p}. \label{51}
\end{eqnarray}
Thus the transverse spin of such a propagating Airy beam can be completely characterized by the Poynting spin. It is the evanescent wave tail characteristics of the Airy beam that lead to similar results to the cosine evanescent wave in Subsec.~\ref{Cosine}. From Fig.~\ref{figs-fig8} (b) - Fig.~\ref{figs-fig8} (d), the spin vectors weigh mainly along the main curves for large $w$ points. In Fig.~\ref{figs-fig8} (e) three types of spins $\tilde{S}_{\rm{p}}$, $\tilde{S}_{\rm{e}}$ and $\tilde{S}$ for $z=2.33\lambda$ are compared as a function of the lateral position $x$, and they obey the double relation Eq.~\eqref{51}, shown from $\delta \tilde{s}_y=0$ in Fig.~\ref{figs-fig8} (e). In Fig.~\ref{figs-fig8} (f) for a line of $z=2.33 \lambda, y=\lambda/2$, the spins gradually decreases with the lateral scale $x$, which demonstrates that evanescent waves appear naturally in Airy beams.

Secondly, for a Hertz vector with $\hat{n}=\hat{z}$, we get similar results with $\hat{n}=\hat{y}$ in Eq.~\eqref{51} (cf. Sec.~$\mathrm{VII}$ of Supplemental Material~\cite{SM}). 

Lastly and the most importantly, when Hertz vector directs where the Airy beam bends, i.e. $\hat{n}=\hat{x}$,
\begin{eqnarray}
\bm{\Pi}_m=A_0Ai(s-\frac{\xi^2}{4})e^{\frac{is\xi}{2}-\frac{i\xi^3}{12}}e^{ikz}\hat{x}, \label{52}
\end{eqnarray}
the extraordinary spin will stand out. Its electromagnetic vector can be written as 
\begin{eqnarray}
\textbf{E}&=&A_0\omega\mu_0e^{i\phi}\left(\begin{array}{ccc}
0\\
-k\mathcal{A}-\frac{x\mathcal{A}}{2kx^3_0}+\frac{z^2\mathcal{A}}{4k^3x^6_0}-\frac{iz\mathcal{A}'}{2k^2x^4_0}\\
0\end{array}\right), \nonumber
\\\label{53}\\
\textbf{H}&=&A_0e^{i\phi}\left(\begin{array}{ccc}
k^2\mathcal{A}+\frac{x\mathcal{A}}{x_0^3}-\frac{z^2\mathcal{A}}{2k^2x^6_0}+\frac{iz\mathcal{A}'}{kx^4_0}\\
\\
0\\
\\
\frac{z^3\mathcal{A}}{4k^4x^9_0}-\frac{3xz\mathcal{A}}{4k^2x^6_0}-\frac{z\mathcal{A}}{2x^3_0}+\frac{i\mathcal{A}}{2kx^3_0}\\
+\frac{ik\mathcal{A}'}{x_0}-\frac{iz^2\mathcal{A}'}{2k^3x^7_0}+\frac{ix\mathcal{A}'}{2kx^4_0}\end{array}\right).\label{54}
\end{eqnarray}
The transverse spin is only provided by the magnetic field in this case. And we calculate the momentum $\textbf{p}$ and SAM $\textbf{s}$ of this Airy field (cf. Sec.~$\mathrm{VII}$ of Supplemental Material~\cite{SM} for the sake of succinctness). 

\begin{figure*}[hbtp!]
  \includegraphics[width=0.94\textwidth]{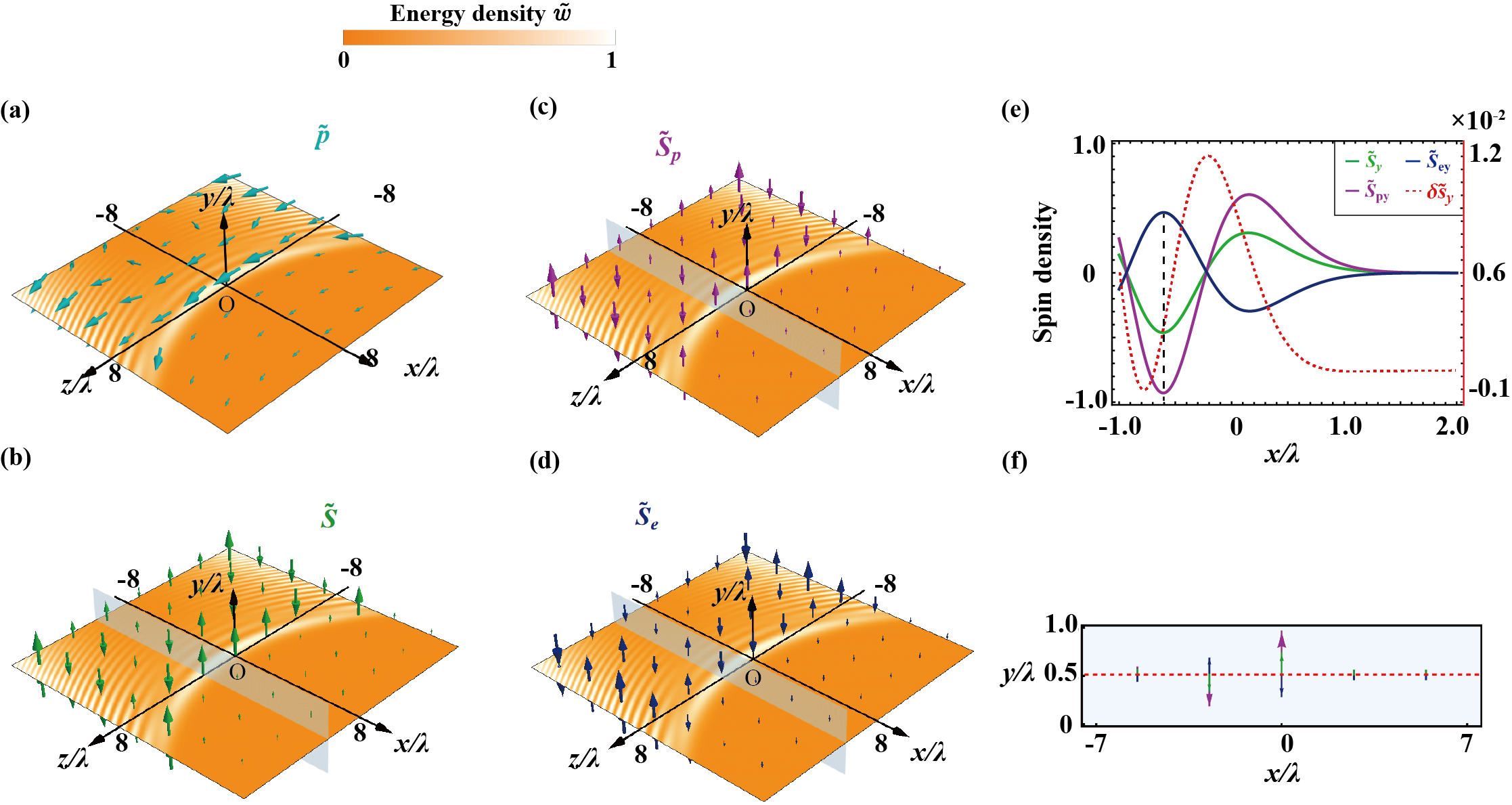}
     \caption{\label{figs-fig9} The Airy beam defined by $\bm{\Pi}_m=\Pi_m\hat{x}$. (a) Normalized momentum $\tilde{p}$ and (b) SAM $\tilde{S}$ vector distribution; (c)-(d) Normalized Poynting spin $\tilde{S}_{\rm{p}}$ and the extraordinary spin $\tilde{S}_{\rm{e}}$ vector distribution of the Airy beam. (e) The three spins density distributions, the black dashed line represents $x=-0.6\lambda$. Here $\delta \tilde{s}_y=\tilde{S}_{\rm{p}y}/2+\tilde{S}_{\rm{e}y}$ reaches its maximum near $x=-0.2\lambda$. (f) the vector distribution of three parts spin at $z=2.33\lambda$ plane. Here, we choose $x_0=0.51\lambda$, the other parameters are the same as Fig.~\ref{figs-fig8}. } 
\end{figure*}

The momentum $\textbf{p}$ for this Airy beam is still distributed in \emph{x-z} propagation plane, and its SAM still completely perpendicular to the propagation plane, similar to the result of Eqs.~\eqref{49} and ~\eqref{50} as show in Figs.~\ref{figs-fig9} (a) and \ref{figs-fig9} (b). Here the spin decomposition of this Airy beam reveals that extraordinary spin just stands out to encapsulate non-planar phase, near fields as well as its elliptic polarization. Using Eqs.~\eqref{12} and ~\eqref{13}, we get $\textbf{s}_{\rm p}$ and $\textbf{s}_{\rm e}$ (cf. Sec.~$\mathrm{VII}$ of Supplemental Material~\cite{SM}). The spin decomposition from the Hertz vector polarizing along its bending direction, in fact disobeys the double relation in Eq.~\eqref{51}. The two parts of spin decomposition are shown respectively in Figs.~\ref{figs-fig9} (c) and \ref{figs-fig9} (d), superimposed on the energy density. And from the three spins densities for plane $z=2.33\lambda$ in Fig.~\ref{figs-fig9} (e), the extraordinary spin $\tilde{S}_e$ breaks the double relation Eq.~\eqref{51} because $\delta \tilde{s}_y=\tilde{S}_{\rm{p}y}/2+\tilde{S}_{\rm{e}y}$ reaches its maximum near $x=-0.2\lambda$ [also see Fig.~\ref{figs-fig9} (f) for the line $z=2.33\lambda, y=0.5\lambda$]. Hence the extraordinary spin steps in as a part of transverse spin to reflect its polarization caused by our free choice of Hertz vector. 

Therefore, the Airy vector beam with its Hertz vector polarizes along the bending direction, reveals an extraordinary spin out of the non-planar phase, near fields as well as its elliptic polarization, which includes every feature resulting to the extraordinary spin, as shown already in the previous three examples. 

\section{Experimental remark and an extension for scalar waves}\label{remarks}
\subsection{Experimental remark}\label{experiment}
We now remark about possible experimental observations for our theoretical framework. For certain surface waves related in our framework, Yin X.'s \cite{YinX2020Spin-resolved, ShiP2021transverse, ShiP2023Dynamical} and Taneja C.'s~\cite{Taneja2021t_spin} experimental setups can directly measure longitudinal component of SAMs, and its transverse component of SAM can be further calculated to reveal the complete vector of SAM. One may also use nonlinear optical technique to retrieve the field information~\cite{Frischwasse2021Real-time}. Therefore it is feasible to adopt such experimental paradigms to measure both near-field and free-space light modes. One calibrates the electric field mapping and calculates the magnetic and Poynting vectors accordingly. Furthermore, such a SAM can be experimentally accessible as a physical quantity measurable from the perspective of angular momenta variance due to torque manipulation. Thus a mechanical measurement of optical force~\cite{ShiY2022stable} and beam shift~\cite{PengL2022spin_Hall} can also reveal its transverse SAM. We also note that such SAM quantity may also be measured by experimental techniques meant for calibrating optical skyrmions~\cite{DuL2019Deep-sub, DaiY2022Ultrafast} and transverse OAM~\cite{ChenW2022Time}.

Setting this stage up to measure SAMs for vector fields, as a possible experimental scheme, one may pinpoint our decomposition by measuring and calculating the aforementioned discrepancy 
\begin{equation}
\delta \tilde{s}=\frac{\tilde{S}_{\rm{p}}}{2}+\tilde{S}_{\rm{e}},
\end{equation}
which should appear significant if our decomposition works, as shown in Figs.~\ref{figs-fig3}, \ref{figs-fig6}, \ref{figs-fig8} and~\ref{figs-fig9}. Though varied in magnitude, one is suggested to look for the tell-tale hallmark of extraordinary spin among them: $\delta \tilde{s}_\phi$ in vector Bessel beam shown in Fig.~\ref{figs-fig3}(e). Here the two normalised $\tilde{S}_{\rm{p}}, \tilde{S}_{\rm{e}}$ are both calculated from the measured electric fields~\cite{YinX2020Spin-resolved, ShiP2023Dynamical}. 
Another possible scheme to test our theory of decomposed SAM, can be to directly calculate the extraordinary spin by subtracting Poynting spin from the whole SAM
\begin{equation}
\mathbf{s}_{\rm e}=\textbf{s}-\textbf{s}_{\rm p}. 
\end{equation}
Therefore our framework is experimentally accessible at least via such schemes to test its applicability experimentally.

 \subsection{An extension for scalar waves}
For theoretical extension, we use the scalar acoustic wave~\cite{Bliokh2019Spin, Bliokh2022Erratum} to exemplify the universal applicability of our theory on SAM decomposition. In such a case, acoustic wave provides no extraordinary SAM at all due to its longitudinal nature~\cite{Bliokh2019t-spin, Vernon2024}. We may define the SAM density and the kinetic momentum respectively for acoustic field phasor $P$ by rotation of velocity field $\mathbf{v}(:=-i\nabla P/ \rho\omega)$
\begin{equation}
\mathbf{s}_{\rm A}= \frac{\rho}{2\omega}\textrm{Im}[\mathbf{v}^*\times \mathbf{v}], 
\end{equation}
and 
\begin{equation}\label{pA}
\mathbf{p}_{\rm A}=\frac{1}{2c_{\rm A}^2}\textrm{Re}[P^*\mathbf{v}]. 
\end{equation}
In Eq.~\eqref{pA} the acoustic velocity $c_{\rm A}$ is defined by reciprocal multiplication of its compressibility $\beta_{\rm A}$ and mass density $\rho$ : $c_{\rm A}^2= 1/\beta_{\rm A}\rho$. With a straightforward algebraic manipulation, one gets 
\begin{equation}
\mathbf{s}_{\rm A}= \frac{c_{\rm A}^2}{\omega^2}\nabla\times \mathbf{p}_{\rm A}-\frac{1}{2\omega^2}\textrm{Re}[P^*\nabla\times \mathbf{v}], 
\end{equation} 
where the latter term on its R. H. S. vanishes due to the longitudinal condition $\nabla\times \mathbf{v}=0$. So acoustic wave fields carry no \emph{extraordinary spin} at all, which is exemplified by acoustic Bessel beam in Sec. VIII of Supplemental Material~\cite{SM}. This null result also justifies the definition of Eq.~\eqref{12} to incorporate the whole curl of kinetic momentum. Hence our decomposition theory in Sec.~\ref{sec:level2} formally applies to generic scalar waves: 
\begin{eqnarray}
\mathbf{s}_{\rm A}&=&\mathbf{s}_{\mathbf{p}}+\mathbf{s}_{\rm e},  \\
\mathbf{s}_{\mathbf{p}}&=& \frac{c_{\rm A}^2}{\omega^2}\nabla\times \mathbf{p}_{\rm A}, \\
\mathbf{s}_{\rm e}&=&-\frac{1}{2\omega^2}\textrm{Re}[P^*\nabla\times \mathbf{v}]. \label{seA}
\end{eqnarray} 
The extraordinary spin in Eq.~\eqref{seA} can well stand out for general scalar waves breaking the longitudinal condition. We note that our decomposition idea shall generally apply to other types of waves~\cite{ShiP2023Dynamical, Vernon2024}, which goes beyond the scope of our paper. However, our formulaic layout for both vector and scalar waves suffices to put alive the principle of dichotomy.

\section{Conclusion}\label{conclusion}

In a vector wave, the connection between its transverse SAM and the curl of momentum was revealed remarkably~\cite{ShiP2021transverse} using the classical fluid description dated back to Maxwell equations, but the complete geometric picture for decomposing its spin angular momenta remains elusive~\cite{Vernon2024}. Along this line of curiosity we are tempted to delve on what makes up the whole SAM, apart from only the curl of momentum. 

Then in this work, using Hertz vector method we analytically build vector beams in general form, and give its explicit decomposition of transverse spins, into the Poynting spin and the extraordinary one. To validate such dichotomy of our framework, concrete examples for transverse SAM of four types of vector beams are applied to clarify the physical origins of transverse spins and to verify its robustness. We find out in these examples that while the Poynting spin derives from the inhomogeneous momentum for vector waves, the newly-defined extraordinary spin is attributed from \emph{the polarization degree-of-freedom, near field, and the non-planar wave front}, which of them is illustrated in the first three types of beams (Subsecs.~\ref{Cosine},~\ref{Bessel} and \ref{Gaussian}) respectively, and all in the last (Subsec.~\ref{AiryBeam}) simultaneously. Moreover, our new decomposition of transverse spins for vector fields applies broadly for both evanescent and propagating waves, and it enriches the concept of transverse spins by adding a new extraordinary part, which results from the higher order derivatives of the structure vector fields. Our paper shall advance the understanding to the spin feature in classical vector waves~\cite{SongC2025Generic}, which applies well to other kinds of spinful scalar wave platform~\cite{Bliokh2022Field, LongY2023}.




\begin{acknowledgments} 

Z.-K. X., Y. L. and M. W. are supported by Young Scientist Fund [NSFC11804087], and Natural National Science Foundation [NSFC62571212, NSFC12075081, NSFC12047501]; Science and Technology Department of Hubei Province [2022CFB553, 2024AFA038]. Y. L. and B. Z. are supported by National Natural Science Foundation of China [NSFC12074107], Science and Technology Department of Hubei Province [2022CFA012], Educational Commission of Hubei Province [T2020001], and the Wuhan city key R\&D program [2025050602030069].  

\end{acknowledgments}

\newpage

\bibliography{Xiong2024Transverse_main_v65}

\providecommand{\noopsort}[1]{}\providecommand{\singleletter}[1]{#1}%
\begin{thebibliography}{66}%
\makeatletter
\providecommand \@ifxundefined [1]{%
 \@ifx{#1\undefined}
}%
\providecommand \@ifnum [1]{%
 \ifnum #1\expandafter \@firstoftwo
 \else \expandafter \@secondoftwo
 \fi
}%
\providecommand \@ifx [1]{%
 \ifx #1\expandafter \@firstoftwo
 \else \expandafter \@secondoftwo
 \fi
}%
\providecommand \natexlab [1]{#1}%
\providecommand \enquote  [1]{``#1''}%
\providecommand \bibnamefont  [1]{#1}%
\providecommand \bibfnamefont [1]{#1}%
\providecommand \citenamefont [1]{#1}%
\providecommand \href@noop [0]{\@secondoftwo}%
\providecommand \href [0]{\begingroup \@sanitize@url \@href}%
\providecommand \@href[1]{\@@startlink{#1}\@@href}%
\providecommand \@@href[1]{\endgroup#1\@@endlink}%
\providecommand \@sanitize@url [0]{\catcode `\\12\catcode `\$12\catcode
  `\&12\catcode `\#12\catcode `\^12\catcode `\_12\catcode `\%12\relax}%
\providecommand \@@startlink[1]{}%
\providecommand \@@endlink[0]{}%
\providecommand \url  [0]{\begingroup\@sanitize@url \@url }%
\providecommand \@url [1]{\endgroup\@href {#1}{\urlprefix }}%
\providecommand \urlprefix  [0]{URL }%
\providecommand \Eprint [0]{\href }%
\providecommand \doibase [0]{https://doi.org/}%
\providecommand \selectlanguage [0]{\@gobble}%
\providecommand \bibinfo  [0]{\@secondoftwo}%
\providecommand \bibfield  [0]{\@secondoftwo}%
\providecommand \translation [1]{[#1]}%
\providecommand \BibitemOpen [0]{}%
\providecommand \bibitemStop [0]{}%
\providecommand \bibitemNoStop [0]{.\EOS\space}%
\providecommand \EOS [0]{\spacefactor3000\relax}%
\providecommand \BibitemShut  [1]{\csname bibitem#1\endcsname}%
\let\auto@bib@innerbib\@empty
\bibitem [{\citenamefont {Jackson}(1999)}]{Jackson1999}%
  \BibitemOpen
  \bibfield  {author} {\bibinfo {author} {\bibfnamefont {J.~D.}\ \bibnamefont
  {Jackson}},\ }\href@noop {} {\emph {\bibinfo {title} {Classical
  Electrodynamics}}},\ \bibinfo {edition} {3rd}\ ed.\ (\bibinfo  {publisher}
  {Wiley},\ \bibinfo {year} {1999})\BibitemShut {NoStop}%
\bibitem [{\citenamefont {Allen}\ \emph {et~al.}(2016)\citenamefont {Allen},
  \citenamefont {Barnett},\ and\ \citenamefont {Padgett}}]{Allen2016}%
  \BibitemOpen
  \bibfield  {author} {\bibinfo {author} {\bibfnamefont {L.}~\bibnamefont
  {Allen}}, \bibinfo {author} {\bibfnamefont {S.~M.}\ \bibnamefont {Barnett}},\
  and\ \bibinfo {author} {\bibfnamefont {M.~J.}\ \bibnamefont {Padgett}},\
  }\href {https://doi.org/10.1201/9781482269017} {\emph {\bibinfo {title}
  {Optical Angular Momentum}}}\ (\bibinfo  {publisher} {CRC press},\ \bibinfo
  {year} {2016})\BibitemShut {NoStop}%
\bibitem [{\citenamefont {Andrews}\ and\ \citenamefont
  {Babiker}(2012)}]{Andrews2012}%
  \BibitemOpen
  \bibfield  {author} {\bibinfo {author} {\bibfnamefont {D.~L.}\ \bibnamefont
  {Andrews}}\ and\ \bibinfo {author} {\bibfnamefont {M.}~\bibnamefont
  {Babiker}},\ }\href {https://doi.org/10.1017/CBO9780511795213} {\emph
  {\bibinfo {title} {The Angular Momentum of Light}}}\ (\bibinfo  {publisher}
  {Cambridge University Press},\ \bibinfo {year} {2012})\BibitemShut {NoStop}%
\bibitem [{\citenamefont {Griffiths}\ and\ \citenamefont
  {Schroeter}(2018)}]{Griffiths2018}%
  \BibitemOpen
  \bibfield  {author} {\bibinfo {author} {\bibfnamefont {D.~J.}\ \bibnamefont
  {Griffiths}}\ and\ \bibinfo {author} {\bibfnamefont {D.~F.}\ \bibnamefont
  {Schroeter}},\ }\href@noop {} {\emph {\bibinfo {title} {Introduction to
  Quantum Mechanics}}}\ (\bibinfo  {publisher} {Cambridge University Press},\
  \bibinfo {year} {2018})\BibitemShut {NoStop}%
\bibitem [{\citenamefont {{\v{Z}}uti{\'c}}\ \emph {et~al.}(2004)\citenamefont
  {{\v{Z}}uti{\'c}}, \citenamefont {Fabian},\ and\ \citenamefont
  {Sarma}}]{Vzutic2004}%
  \BibitemOpen
  \bibfield  {author} {\bibinfo {author} {\bibfnamefont {I.}~\bibnamefont
  {{\v{Z}}uti{\'c}}}, \bibinfo {author} {\bibfnamefont {J.}~\bibnamefont
  {Fabian}},\ and\ \bibinfo {author} {\bibfnamefont {S.~D.}\ \bibnamefont
  {Sarma}},\ }\bibfield  {title} {\bibinfo {title} {Spintronics: Fundamentals
  and applications},\ }\href {https://doi.org/10.1103/RevModPhys.76.323}
  {\bibfield  {journal} {\bibinfo  {journal} {Reviews of {M}odern {P}hysics}\
  }\textbf {\bibinfo {volume} {76}},\ \bibinfo {pages} {323} (\bibinfo {year}
  {2004})}\BibitemShut {NoStop}%
\bibitem [{\citenamefont {Griffiths}(2013)}]{Griffiths2013}%
  \BibitemOpen
  \bibfield  {author} {\bibinfo {author} {\bibfnamefont {D.~J.}\ \bibnamefont
  {Griffiths}},\ }\href@noop {} {\emph {\bibinfo {title} {Introduction to
  Electrodynamics}}}\ (\bibinfo  {publisher} {Cambridge University Press},\
  \bibinfo {year} {2013})\BibitemShut {NoStop}%
\bibitem [{\citenamefont {Ishimaru}(2017)}]{Ishimaru2017}%
  \BibitemOpen
  \bibfield  {author} {\bibinfo {author} {\bibfnamefont {A.}~\bibnamefont
  {Ishimaru}},\ }\href {DOI:10.1002/9781119079699} {\emph {\bibinfo {title}
  {Electromagnetic wave propagation, radiation, and scattering: from
  fundamentals to applications}}}\ (\bibinfo  {publisher} {John Wiley \&
  Sons},\ \bibinfo {year} {2017})\BibitemShut {NoStop}%
\bibitem [{\citenamefont {Bliokh}\ \emph
  {et~al.}(2015{\natexlab{a}})\citenamefont {Bliokh}, \citenamefont
  {Rodr{\'\i}guez-Fortu{\~n}o}, \citenamefont {Nori},\ and\ \citenamefont
  {Zayats}}]{Bliokh2015spin}%
  \BibitemOpen
  \bibfield  {author} {\bibinfo {author} {\bibfnamefont {K.~Y.}\ \bibnamefont
  {Bliokh}}, \bibinfo {author} {\bibfnamefont {F.~J.}\ \bibnamefont
  {Rodr{\'\i}guez-Fortu{\~n}o}}, \bibinfo {author} {\bibfnamefont
  {F.}~\bibnamefont {Nori}},\ and\ \bibinfo {author} {\bibfnamefont {A.~V.}\
  \bibnamefont {Zayats}},\ }\bibfield  {title} {\bibinfo {title} {Spin-orbit
  interactions of light},\ }\href {https://doi.org/10.1038/nphoton.2015.201}
  {\bibfield  {journal} {\bibinfo  {journal} {Nature Photonics}\ }\textbf
  {\bibinfo {volume} {9}},\ \bibinfo {pages} {796} (\bibinfo {year}
  {2015}{\natexlab{a}})}\BibitemShut {NoStop}%
\bibitem [{\citenamefont {Bliokh}\ and\ \citenamefont
  {Nori}(2012)}]{Bliokh2012}%
  \BibitemOpen
  \bibfield  {author} {\bibinfo {author} {\bibfnamefont {K.~Y.}\ \bibnamefont
  {Bliokh}}\ and\ \bibinfo {author} {\bibfnamefont {F.}~\bibnamefont {Nori}},\
  }\bibfield  {title} {\bibinfo {title} {Transverse spin of a surface
  polariton},\ }\href@noop {} {\bibfield  {journal} {\bibinfo  {journal}
  {Physical Review A}\ }\textbf {\bibinfo {volume} {85}},\ \bibinfo {pages}
  {061801} (\bibinfo {year} {2012})}\BibitemShut {NoStop}%
\bibitem [{\citenamefont {Bliokh}\ \emph {et~al.}(2014)\citenamefont {Bliokh},
  \citenamefont {Bekshaev},\ and\ \citenamefont {Nori}}]{Bliokh2014}%
  \BibitemOpen
  \bibfield  {author} {\bibinfo {author} {\bibfnamefont {K.~Y.}\ \bibnamefont
  {Bliokh}}, \bibinfo {author} {\bibfnamefont {A.~Y.}\ \bibnamefont
  {Bekshaev}},\ and\ \bibinfo {author} {\bibfnamefont {F.}~\bibnamefont
  {Nori}},\ }\bibfield  {title} {\bibinfo {title} {Extraordinary momentum and
  spin in evanescent waves},\ }\href {https://doi.org/10.1038/ncomms4300}
  {\bibfield  {journal} {\bibinfo  {journal} {Nature Communications}\ }\textbf
  {\bibinfo {volume} {5}},\ \bibinfo {pages} {3300} (\bibinfo {year}
  {2014})}\BibitemShut {NoStop}%
\bibitem [{\citenamefont {Bekshaev}\ \emph {et~al.}(2015)\citenamefont
  {Bekshaev}, \citenamefont {Bliokh},\ and\ \citenamefont
  {Nori}}]{Bekshaev2015}%
  \BibitemOpen
  \bibfield  {author} {\bibinfo {author} {\bibfnamefont {A.~Y.}\ \bibnamefont
  {Bekshaev}}, \bibinfo {author} {\bibfnamefont {K.~Y.}\ \bibnamefont
  {Bliokh}},\ and\ \bibinfo {author} {\bibfnamefont {F.}~\bibnamefont {Nori}},\
  }\bibfield  {title} {\bibinfo {title} {Transverse spin and momentum in
  two-wave interference},\ }\href
  {https://journals.aps.org/prx/pdf/10.1103/PhysRevX.5.011039} {\bibfield
  {journal} {\bibinfo  {journal} {Physical Review X}\ }\textbf {\bibinfo
  {volume} {5}},\ \bibinfo {pages} {011039} (\bibinfo {year}
  {2015})}\BibitemShut {NoStop}%
\bibitem [{\citenamefont {Aiello}\ and\ \citenamefont
  {Banzer}(2015)}]{Aiello2015transverse}%
  \BibitemOpen
  \bibfield  {author} {\bibinfo {author} {\bibfnamefont {A.}~\bibnamefont
  {Aiello}}\ and\ \bibinfo {author} {\bibfnamefont {P.}~\bibnamefont
  {Banzer}},\ }\bibfield  {title} {\bibinfo {title} {Transverse spin of light
  for all wavefields},\ }\href {https://doi.org/10.1038/nphoton.2015.203}
  {\bibfield  {journal} {\bibinfo  {journal} {arXiv preprint arXiv:1502.05350}\
  } (\bibinfo {year} {2015})}\BibitemShut {NoStop}%
\bibitem [{\citenamefont {Bliokh}\ and\ \citenamefont
  {Nori}(2015)}]{Bliokh2015Transverse}%
  \BibitemOpen
  \bibfield  {author} {\bibinfo {author} {\bibfnamefont {K.~Y.}\ \bibnamefont
  {Bliokh}}\ and\ \bibinfo {author} {\bibfnamefont {F.}~\bibnamefont {Nori}},\
  }\bibfield  {title} {\bibinfo {title} {Transverse and longitudinal angular
  momenta of light},\ }\href {https://doi.org/10.1016/j.physrep.2015.06.003}
  {\bibfield  {journal} {\bibinfo  {journal} {Physics Reports}\ }\textbf
  {\bibinfo {volume} {592}},\ \bibinfo {pages} {1} (\bibinfo {year}
  {2015})}\BibitemShut {NoStop}%
\bibitem [{\citenamefont {Aiello}\ \emph {et~al.}(2015)\citenamefont {Aiello},
  \citenamefont {Banzer}, \citenamefont {Neugebauer},\ and\ \citenamefont
  {Leuchs}}]{Aiello2015}%
  \BibitemOpen
  \bibfield  {author} {\bibinfo {author} {\bibfnamefont {A.}~\bibnamefont
  {Aiello}}, \bibinfo {author} {\bibfnamefont {P.}~\bibnamefont {Banzer}},
  \bibinfo {author} {\bibfnamefont {M.}~\bibnamefont {Neugebauer}},\ and\
  \bibinfo {author} {\bibfnamefont {G.}~\bibnamefont {Leuchs}},\ }\bibfield
  {title} {\bibinfo {title} {From transverse angular momentum to photonic
  wheels},\ }\href {https://www.nature.com/articles/nphoton.2015.203}
  {\bibfield  {journal} {\bibinfo  {journal} {Nature Photonics}\ }\textbf
  {\bibinfo {volume} {9}},\ \bibinfo {pages} {789} (\bibinfo {year}
  {2015})}\BibitemShut {NoStop}%
\bibitem [{\citenamefont {O’connor}\ \emph {et~al.}(2014)\citenamefont
  {O’connor}, \citenamefont {Ginzburg}, \citenamefont
  {Rodr{\'\i}guez-Fortu{\~n}o}, \citenamefont {Wurtz},\ and\ \citenamefont
  {Zayats}}]{Oconnor2014}%
  \BibitemOpen
  \bibfield  {author} {\bibinfo {author} {\bibfnamefont {D.}~\bibnamefont
  {O’connor}}, \bibinfo {author} {\bibfnamefont {P.}~\bibnamefont
  {Ginzburg}}, \bibinfo {author} {\bibfnamefont {F.~J.}\ \bibnamefont
  {Rodr{\'\i}guez-Fortu{\~n}o}}, \bibinfo {author} {\bibfnamefont {G.~A.}\
  \bibnamefont {Wurtz}},\ and\ \bibinfo {author} {\bibfnamefont {A.~V.}\
  \bibnamefont {Zayats}},\ }\bibfield  {title} {\bibinfo {title} {Spin--orbit
  coupling in surface plasmon scattering by nanostructures},\ }\href
  {https://www.nature.com/articles/ncomms6327} {\bibfield  {journal} {\bibinfo
  {journal} {Nature Communications}\ }\textbf {\bibinfo {volume} {5}},\
  \bibinfo {pages} {5327} (\bibinfo {year} {2014})}\BibitemShut {NoStop}%
\bibitem [{\citenamefont {Mitsch}\ \emph {et~al.}(2014)\citenamefont {Mitsch},
  \citenamefont {Sayrin}, \citenamefont {Albrecht}, \citenamefont
  {Schneeweiss},\ and\ \citenamefont {Rauschenbeutel}}]{Mitsch2014}%
  \BibitemOpen
  \bibfield  {author} {\bibinfo {author} {\bibfnamefont {R.}~\bibnamefont
  {Mitsch}}, \bibinfo {author} {\bibfnamefont {C.}~\bibnamefont {Sayrin}},
  \bibinfo {author} {\bibfnamefont {B.}~\bibnamefont {Albrecht}}, \bibinfo
  {author} {\bibfnamefont {P.}~\bibnamefont {Schneeweiss}},\ and\ \bibinfo
  {author} {\bibfnamefont {A.}~\bibnamefont {Rauschenbeutel}},\ }\bibfield
  {title} {\bibinfo {title} {Quantum state-controlled directional spontaneous
  emission of photons into a nanophotonic waveguide},\ }\href
  {https://doi.org/10.1038/ncomms6713} {\bibfield  {journal} {\bibinfo
  {journal} {Nature Communications}\ }\textbf {\bibinfo {volume} {5}},\
  \bibinfo {pages} {5713} (\bibinfo {year} {2014})}\BibitemShut {NoStop}%
\bibitem [{\citenamefont {Rodríguez-Fortuño}\ \emph
  {et~al.}(2013)\citenamefont {Rodríguez-Fortuño}, \citenamefont {Marino},
  \citenamefont {Ginzburg}, \citenamefont {O’Connor}, \citenamefont
  {Martínez}, \citenamefont {Wurtz},\ and\ \citenamefont
  {Zayats}}]{Francisco2013}%
  \BibitemOpen
  \bibfield  {author} {\bibinfo {author} {\bibfnamefont {F.~J.}\ \bibnamefont
  {Rodríguez-Fortuño}}, \bibinfo {author} {\bibfnamefont {G.}~\bibnamefont
  {Marino}}, \bibinfo {author} {\bibfnamefont {P.}~\bibnamefont {Ginzburg}},
  \bibinfo {author} {\bibfnamefont {D.}~\bibnamefont {O’Connor}}, \bibinfo
  {author} {\bibfnamefont {A.}~\bibnamefont {Martínez}}, \bibinfo {author}
  {\bibfnamefont {G.~A.}\ \bibnamefont {Wurtz}},\ and\ \bibinfo {author}
  {\bibfnamefont {A.~V.}\ \bibnamefont {Zayats}},\ }\bibfield  {title}
  {\bibinfo {title} {Near-field interference for the unidirectional excitation
  of electromagnetic guided modes},\ }\href
  {https://doi.org/10.1126/science.1233739} {\bibfield  {journal} {\bibinfo
  {journal} {Science}\ }\textbf {\bibinfo {volume} {340}},\ \bibinfo {pages}
  {328} (\bibinfo {year} {2013})}\BibitemShut {NoStop}%
\bibitem [{\citenamefont {Petersen}\ \emph {et~al.}(2014)\citenamefont
  {Petersen}, \citenamefont {Volz},\ and\ \citenamefont
  {Rauschenbeutel}}]{Petersen2014}%
  \BibitemOpen
  \bibfield  {author} {\bibinfo {author} {\bibfnamefont {J.}~\bibnamefont
  {Petersen}}, \bibinfo {author} {\bibfnamefont {J.}~\bibnamefont {Volz}},\
  and\ \bibinfo {author} {\bibfnamefont {A.}~\bibnamefont {Rauschenbeutel}},\
  }\bibfield  {title} {\bibinfo {title} {Chiral nanophotonic waveguide
  interface based on spin-orbit interaction of light},\ }\href
  {https://www.science.org/doi/abs/10.1126/science.1257671} {\bibfield
  {journal} {\bibinfo  {journal} {Science}\ }\textbf {\bibinfo {volume}
  {346}},\ \bibinfo {pages} {67} (\bibinfo {year} {2014})}\BibitemShut
  {NoStop}%
\bibitem [{\citenamefont {Deng}\ \emph {et~al.}(2017)\citenamefont {Deng},
  \citenamefont {Chen}, \citenamefont {Zhao},\ and\ \citenamefont
  {Dong}}]{Deng2017}%
  \BibitemOpen
  \bibfield  {author} {\bibinfo {author} {\bibfnamefont {W.-M.}\ \bibnamefont
  {Deng}}, \bibinfo {author} {\bibfnamefont {X.-D.}\ \bibnamefont {Chen}},
  \bibinfo {author} {\bibfnamefont {F.-L.}\ \bibnamefont {Zhao}},\ and\
  \bibinfo {author} {\bibfnamefont {J.-W.}\ \bibnamefont {Dong}},\ }\bibfield
  {title} {\bibinfo {title} {Transverse angular momentum in topological
  photonic crystals},\ }\href
  {https://iopscience.iop.org/article/10.1088/2040-8986/aa9b06/meta} {\bibfield
   {journal} {\bibinfo  {journal} {Journal of Optics}\ }\textbf {\bibinfo
  {volume} {20}},\ \bibinfo {pages} {014006} (\bibinfo {year}
  {2017})}\BibitemShut {NoStop}%
\bibitem [{\citenamefont {Pal}\ \emph {et~al.}(2020)\citenamefont {Pal},
  \citenamefont {Gupta}, \citenamefont {Ghosh},\ and\ \citenamefont
  {Banerjee}}]{Pal2020}%
  \BibitemOpen
  \bibfield  {author} {\bibinfo {author} {\bibfnamefont {D.}~\bibnamefont
  {Pal}}, \bibinfo {author} {\bibfnamefont {S.~D.}\ \bibnamefont {Gupta}},
  \bibinfo {author} {\bibfnamefont {N.}~\bibnamefont {Ghosh}},\ and\ \bibinfo
  {author} {\bibfnamefont {A.}~\bibnamefont {Banerjee}},\ }\bibfield  {title}
  {\bibinfo {title} {Direct observation of the effects of spin dependent
  momentum of light in optical tweezers},\ }\href
  {https://doi.org/10.1063/5.0015991} {\bibfield  {journal} {\bibinfo
  {journal} {APL Photonics}\ }\textbf {\bibinfo {volume} {5}} (\bibinfo {year}
  {2020})}\BibitemShut {NoStop}%
\bibitem [{\citenamefont {Shi}\ \emph {et~al.}(2023{\natexlab{a}})\citenamefont
  {Shi}, \citenamefont {Xu}, \citenamefont {Qiu},\ and\ \citenamefont
  {Cheng}}]{ShiY2023}%
  \BibitemOpen
  \bibfield  {author} {\bibinfo {author} {\bibfnamefont {Y.}~\bibnamefont
  {Shi}}, \bibinfo {author} {\bibfnamefont {X.}~\bibnamefont {Xu}}, \bibinfo
  {author} {\bibfnamefont {C.-W.}\ \bibnamefont {Qiu}},\ and\ \bibinfo {author}
  {\bibfnamefont {X.~e.~a.}\ \bibnamefont {Cheng}},\ }\bibfield  {title}
  {\bibinfo {title} {Advances in light transverse momenta and optical lateral
  forces},\ }\href {https://doi.org/10.1364/AOP.489300} {\bibfield  {journal}
  {\bibinfo  {journal} {Adv. Opt. Photon.}\ }\textbf {\bibinfo {volume} {15}},\
  \bibinfo {pages} {835} (\bibinfo {year} {2023}{\natexlab{a}})}\BibitemShut
  {NoStop}%
\bibitem [{\citenamefont {Van~Mechelen}\ and\ \citenamefont
  {Jacob}(2016)}]{Van2016}%
  \BibitemOpen
  \bibfield  {author} {\bibinfo {author} {\bibfnamefont {T.}~\bibnamefont
  {Van~Mechelen}}\ and\ \bibinfo {author} {\bibfnamefont {Z.}~\bibnamefont
  {Jacob}},\ }\bibfield  {title} {\bibinfo {title} {Universal spin-momentum
  locking of evanescent waves},\ }\href
  {https://doi.org/10.1364/OPTICA.3.000118} {\bibfield  {journal} {\bibinfo
  {journal} {Optica}\ }\textbf {\bibinfo {volume} {3}},\ \bibinfo {pages} {118}
  (\bibinfo {year} {2016})}\BibitemShut {NoStop}%
\bibitem [{\citenamefont {Bliokh}\ \emph
  {et~al.}(2015{\natexlab{b}})\citenamefont {Bliokh}, \citenamefont
  {Smirnova},\ and\ \citenamefont {Nori}}]{Bliokh2015quantum}%
  \BibitemOpen
  \bibfield  {author} {\bibinfo {author} {\bibfnamefont {K.~Y.}\ \bibnamefont
  {Bliokh}}, \bibinfo {author} {\bibfnamefont {D.}~\bibnamefont {Smirnova}},\
  and\ \bibinfo {author} {\bibfnamefont {F.}~\bibnamefont {Nori}},\ }\bibfield
  {title} {\bibinfo {title} {Quantum spin hall effect of light},\ }\href
  {https://www.science.org/doi/abs/10.1126/science.aaa9519} {\bibfield
  {journal} {\bibinfo  {journal} {Science}\ }\textbf {\bibinfo {volume}
  {348}},\ \bibinfo {pages} {1448} (\bibinfo {year}
  {2015}{\natexlab{b}})}\BibitemShut {NoStop}%
\bibitem [{\citenamefont {Shi}\ \emph {et~al.}(2023{\natexlab{b}})\citenamefont
  {Shi}, \citenamefont {Du}, \citenamefont {Yang}, \citenamefont {Yin},
  \citenamefont {Lei},\ and\ \citenamefont {Yuan}}]{ShiP2023Dynamical}%
  \BibitemOpen
  \bibfield  {author} {\bibinfo {author} {\bibfnamefont {P.}~\bibnamefont
  {Shi}}, \bibinfo {author} {\bibfnamefont {L.}~\bibnamefont {Du}}, \bibinfo
  {author} {\bibfnamefont {A.}~\bibnamefont {Yang}}, \bibinfo {author}
  {\bibfnamefont {X.}~\bibnamefont {Yin}}, \bibinfo {author} {\bibfnamefont
  {X.}~\bibnamefont {Lei}},\ and\ \bibinfo {author} {\bibfnamefont
  {X.}~\bibnamefont {Yuan}},\ }\bibfield  {title} {\bibinfo {title} {Dynamical
  and topological properties of the spin angular momenta in general
  electromagnetic fields},\ }\href {https://doi.org/10.1038/s42005-023-01374-y}
  {\bibfield  {journal} {\bibinfo  {journal} {Communications Physics}\ }\textbf
  {\bibinfo {volume} {6}},\ \bibinfo {pages} {283} (\bibinfo {year}
  {2023}{\natexlab{b}})}\BibitemShut {NoStop}%
\bibitem [{\citenamefont {Le~Feber}\ \emph {et~al.}(2015)\citenamefont
  {Le~Feber}, \citenamefont {Rotenberg},\ and\ \citenamefont
  {Kuipers}}]{Le2015}%
  \BibitemOpen
  \bibfield  {author} {\bibinfo {author} {\bibfnamefont {B.}~\bibnamefont
  {Le~Feber}}, \bibinfo {author} {\bibfnamefont {N.}~\bibnamefont
  {Rotenberg}},\ and\ \bibinfo {author} {\bibfnamefont {L.}~\bibnamefont
  {Kuipers}},\ }\bibfield  {title} {\bibinfo {title} {Nanophotonic control of
  circular dipole emission},\ }\href
  {https://www.nature.com/articles/ncomms7695} {\bibfield  {journal} {\bibinfo
  {journal} {Nature Communications}\ }\textbf {\bibinfo {volume} {6}},\
  \bibinfo {pages} {6695} (\bibinfo {year} {2015})}\BibitemShut {NoStop}%
\bibitem [{\citenamefont {Lu}\ \emph {et~al.}(2014)\citenamefont {Lu},
  \citenamefont {Joannopoulos},\ and\ \citenamefont
  {Solja{\v{c}}i{\'c}}}]{Lu2014}%
  \BibitemOpen
  \bibfield  {author} {\bibinfo {author} {\bibfnamefont {L.}~\bibnamefont
  {Lu}}, \bibinfo {author} {\bibfnamefont {J.~D.}\ \bibnamefont
  {Joannopoulos}},\ and\ \bibinfo {author} {\bibfnamefont {M.}~\bibnamefont
  {Solja{\v{c}}i{\'c}}},\ }\bibfield  {title} {\bibinfo {title} {Topological
  photonics},\ }\href {https://www.nature.com/articles/nphoton.2014.248}
  {\bibfield  {journal} {\bibinfo  {journal} {Nature Photonics}\ }\textbf
  {\bibinfo {volume} {8}},\ \bibinfo {pages} {821} (\bibinfo {year}
  {2014})}\BibitemShut {NoStop}%
\bibitem [{\citenamefont {Gustafsson}(2000)}]{Gustafsson2000}%
  \BibitemOpen
  \bibfield  {author} {\bibinfo {author} {\bibfnamefont {M.~G.}\ \bibnamefont
  {Gustafsson}},\ }\bibfield  {title} {\bibinfo {title} {Surpassing the lateral
  resolution limit by a factor of two using structured illumination
  microscopy},\ }\href {https://doi.org/10.1046/j.1365-2818.2000.00710.x}
  {\bibfield  {journal} {\bibinfo  {journal} {Journal of microscopy}\ }\textbf
  {\bibinfo {volume} {198}},\ \bibinfo {pages} {82} (\bibinfo {year}
  {2000})}\BibitemShut {NoStop}%
\bibitem [{\citenamefont {Shi}\ \emph {et~al.}(2021)\citenamefont {Shi},
  \citenamefont {Du}, \citenamefont {Li}, \citenamefont {Zayats},\ and\
  \citenamefont {Yuan}}]{ShiP2021transverse}%
  \BibitemOpen
  \bibfield  {author} {\bibinfo {author} {\bibfnamefont {P.}~\bibnamefont
  {Shi}}, \bibinfo {author} {\bibfnamefont {L.}~\bibnamefont {Du}}, \bibinfo
  {author} {\bibfnamefont {C.}~\bibnamefont {Li}}, \bibinfo {author}
  {\bibfnamefont {A.~V.}\ \bibnamefont {Zayats}},\ and\ \bibinfo {author}
  {\bibfnamefont {X.}~\bibnamefont {Yuan}},\ }\bibfield  {title} {\bibinfo
  {title} {Transverse spin dynamics in structured electromagnetic guided
  waves},\ }\href {https://doi.org/10.1073/pnas.2018816118} {\bibfield
  {journal} {\bibinfo  {journal} {Proceedings of the National Academy of
  Sciences}\ }\textbf {\bibinfo {volume} {118}},\ \bibinfo {pages}
  {e2018816118} (\bibinfo {year} {2021})}\BibitemShut {NoStop}%
\bibitem [{\citenamefont {Shi}\ \emph {et~al.}(2022{\natexlab{a}})\citenamefont
  {Shi}, \citenamefont {Li}, \citenamefont {Du},\ and\ \citenamefont
  {Yuan}}]{ShiP2022}%
  \BibitemOpen
  \bibfield  {author} {\bibinfo {author} {\bibfnamefont {P.}~\bibnamefont
  {Shi}}, \bibinfo {author} {\bibfnamefont {H.}~\bibnamefont {Li}}, \bibinfo
  {author} {\bibfnamefont {L.}~\bibnamefont {Du}},\ and\ \bibinfo {author}
  {\bibfnamefont {X.}~\bibnamefont {Yuan}},\ }\bibfield  {title} {\bibinfo
  {title} {Spin-momentum properties in the paraxial optical systems},\ }\href
  {https://doi.org/10.1021/acsphotonics.2c01535} {\bibfield  {journal}
  {\bibinfo  {journal} {ACS Photonics}\ } (\bibinfo {year}
  {2022}{\natexlab{a}})}\BibitemShut {NoStop}%
\bibitem [{\citenamefont {Song}\ \emph {et~al.}(2025)\citenamefont {Song},
  \citenamefont {He},\ and\ \citenamefont {Yuan}}]{SongC2025Generic}%
  \BibitemOpen
  \bibfield  {author} {\bibinfo {author} {\bibfnamefont {C.}~\bibnamefont
  {Song}}, \bibinfo {author} {\bibfnamefont {J.}~\bibnamefont {He}},\ and\
  \bibinfo {author} {\bibfnamefont {G.}~\bibnamefont {Yuan}},\ }\bibfield
  {title} {\bibinfo {title} {Generic full-vector angular spectrum method for
  calculating diffraction of arbitrary electromagnetic fields},\ }\bibfield
  {journal} {\bibinfo  {journal} {Journal of Physics: Photonics}\ }\textbf
  {\bibinfo {volume} {7}},\ \href {https://doi.org/10.1088/2515-7647/ae0384}
  {10.1088/2515-7647/ae0384} (\bibinfo {year} {2025})\BibitemShut {NoStop}%
\bibitem [{\citenamefont {Bekshaev}(2022)}]{Bekshaev2022t-spin}%
  \BibitemOpen
  \bibfield  {author} {\bibinfo {author} {\bibfnamefont {A.~Y.}\ \bibnamefont
  {Bekshaev}},\ }\bibfield  {title} {\bibinfo {title} {Transverse spin and the
  hidden vorticity of propagating light fields},\ }\href
  {https://doi.org/10.1364/JOSAA.466360} {\bibfield  {journal} {\bibinfo
  {journal} {J Opt Soc Am A Opt Image Sci Vis}\ }\textbf {\bibinfo {volume}
  {39}},\ \bibinfo {pages} {1577} (\bibinfo {year} {2022})}\BibitemShut
  {NoStop}%
\bibitem [{\citenamefont {Vernon}\ \emph {et~al.}(2024)\citenamefont {Vernon},
  \citenamefont {Golat}, \citenamefont {Rigouzzo}, \citenamefont {Lim},\ and\
  \citenamefont {Rodr{\'\i}guez-Fortu{\~n}o}}]{Vernon2024}%
  \BibitemOpen
  \bibfield  {author} {\bibinfo {author} {\bibfnamefont {A.~J.}\ \bibnamefont
  {Vernon}}, \bibinfo {author} {\bibfnamefont {S.}~\bibnamefont {Golat}},
  \bibinfo {author} {\bibfnamefont {C.}~\bibnamefont {Rigouzzo}}, \bibinfo
  {author} {\bibfnamefont {E.~A.}\ \bibnamefont {Lim}},\ and\ \bibinfo {author}
  {\bibfnamefont {F.~J.}\ \bibnamefont {Rodr{\'\i}guez-Fortu{\~n}o}},\
  }\bibfield  {title} {\bibinfo {title} {A decomposition of light’s spin
  angular momentum density},\ }\href
  {https://doi.org/10.1038/s41377-024-01447-9} {\bibfield  {journal} {\bibinfo
  {journal} {Light: Science \& Applications}\ }\textbf {\bibinfo {volume}
  {13}},\ \bibinfo {pages} {160} (\bibinfo {year} {2024})}\BibitemShut
  {NoStop}%
\bibitem [{\citenamefont {Essex}(1977)}]{Essex1977}%
  \BibitemOpen
  \bibfield  {author} {\bibinfo {author} {\bibfnamefont {E.}~\bibnamefont
  {Essex}},\ }\bibfield  {title} {\bibinfo {title} {Hertz vector potentials of
  electromagnetic theory},\ }\href {https://doi.org/10.1119/1.10955} {\bibfield
   {journal} {\bibinfo  {journal} {American Journal of Physics}\ }\textbf
  {\bibinfo {volume} {45}},\ \bibinfo {pages} {1099} (\bibinfo {year}
  {1977})}\BibitemShut {NoStop}%
\bibitem [{\citenamefont {Stratton}(2007)}]{Stratton2007}%
  \BibitemOpen
  \bibfield  {author} {\bibinfo {author} {\bibfnamefont {J.~A.}\ \bibnamefont
  {Stratton}},\ }\href@noop {} {\emph {\bibinfo {title} {Electromagnetic
  Theory}}},\ Vol.~\bibinfo {volume} {33}\ (\bibinfo  {publisher} {John Wiley
  \& Sons},\ \bibinfo {year} {2007})\BibitemShut {NoStop}%
\bibitem [{\citenamefont {Ornigotti}\ and\ \citenamefont
  {Aiello}(2014)}]{Ornigotti2014}%
  \BibitemOpen
  \bibfield  {author} {\bibinfo {author} {\bibfnamefont {M.}~\bibnamefont
  {Ornigotti}}\ and\ \bibinfo {author} {\bibfnamefont {A.}~\bibnamefont
  {Aiello}},\ }\bibfield  {title} {\bibinfo {title} {The hertz vector
  revisited: a simple physical picture},\ }\href
  {https://iopscience.iop.org/article/10.1088/2040-8978/16/10/105705/meta}
  {\bibfield  {journal} {\bibinfo  {journal} {Journal of Optics}\ }\textbf
  {\bibinfo {volume} {16}},\ \bibinfo {pages} {105705} (\bibinfo {year}
  {2014})}\BibitemShut {NoStop}%
\bibitem [{\citenamefont {Wang}\ \emph {et~al.}(2014)\citenamefont {Wang},
  \citenamefont {Dou},\ and\ \citenamefont {Meng}}]{Wang2014}%
  \BibitemOpen
  \bibfield  {author} {\bibinfo {author} {\bibfnamefont {Y.}~\bibnamefont
  {Wang}}, \bibinfo {author} {\bibfnamefont {W.}~\bibnamefont {Dou}},\ and\
  \bibinfo {author} {\bibfnamefont {H.}~\bibnamefont {Meng}},\ }\bibfield
  {title} {\bibinfo {title} {Vector analyses of linearly and circularly
  polarized bessel beams using hertz vector potentials},\ }\href
  {https://doi.org/10.1364/OE.22.007821} {\bibfield  {journal} {\bibinfo
  {journal} {Optics Express}\ }\textbf {\bibinfo {volume} {22}},\ \bibinfo
  {pages} {7821} (\bibinfo {year} {2014})}\BibitemShut {NoStop}%
\bibitem [{\citenamefont {Berry}(2009)}]{Berry2009Opt_currents}%
  \BibitemOpen
  \bibfield  {author} {\bibinfo {author} {\bibfnamefont {M.~V.}\ \bibnamefont
  {Berry}},\ }\bibfield  {title} {\bibinfo {title} {Optical currents},\ }\href
  {https://iopscience.iop.org/article/10.1088/1464-4258/11/9/094001/meta}
  {\bibfield  {journal} {\bibinfo  {journal} {Journal of Optics A: Pure and
  Applied Optics}\ }\textbf {\bibinfo {volume} {11}},\ \bibinfo {pages}
  {094001} (\bibinfo {year} {2009})}\BibitemShut {NoStop}%
\bibitem [{\citenamefont {Bekshaev}(2013)}]{Bekshaev2013Sub}%
  \BibitemOpen
  \bibfield  {author} {\bibinfo {author} {\bibfnamefont {A.~Y.}\ \bibnamefont
  {Bekshaev}},\ }\bibfield  {title} {\bibinfo {title} {Subwavelength particles
  in an inhomogeneous light field: optical forces associated with the spin and
  orbital energy flows},\ }\bibfield  {journal} {\bibinfo  {journal} {Journal
  of Optics}\ }\textbf {\bibinfo {volume} {15}},\ \href
  {https://doi.org/10.1088/2040-8978/15/4/044004}
  {10.1088/2040-8978/15/4/044004} (\bibinfo {year} {2013})\BibitemShut
  {NoStop}%
\bibitem [{\citenamefont {Bliokh}\ \emph {et~al.}(2016)\citenamefont {Bliokh},
  \citenamefont {Bekshaev},\ and\ \citenamefont {Nori}}]{Bliokh2016}%
  \BibitemOpen
  \bibfield  {author} {\bibinfo {author} {\bibfnamefont {K.~Y.}\ \bibnamefont
  {Bliokh}}, \bibinfo {author} {\bibfnamefont {A.~Y.}\ \bibnamefont
  {Bekshaev}},\ and\ \bibinfo {author} {\bibfnamefont {F.}~\bibnamefont
  {Nori}},\ }\bibfield  {title} {\bibinfo {title} {Corrigendum: dual
  electromagnetism: helicity, spin, momentum, and angular momentum (2013 new j.
  phys. 15 033026)},\ }\href
  {https://s3.cern.ch/inspire-prod-files-3/3203da15713d7c8db4522724fb706cc9}
  {\bibfield  {journal} {\bibinfo  {journal} {New Journal of Physics}\ }\textbf
  {\bibinfo {volume} {18}},\ \bibinfo {pages} {089503} (\bibinfo {year}
  {2016})}\BibitemShut {NoStop}%
\bibitem [{\citenamefont {Mao}\ \emph {et~al.}(2021)\citenamefont {Mao},
  \citenamefont {Liu},\ and\ \citenamefont {Lin}}]{MaoY2021}%
  \BibitemOpen
  \bibfield  {author} {\bibinfo {author} {\bibfnamefont {Y.}~\bibnamefont
  {Mao}}, \bibinfo {author} {\bibfnamefont {Y.}~\bibnamefont {Liu}},\ and\
  \bibinfo {author} {\bibfnamefont {H.}~\bibnamefont {Lin}},\ }\bibfield
  {title} {\bibinfo {title} {Angular momenta in fields from a rotational
  mechanical antenna},\ }\href {https://doi.org/10.1088/2399-6528/ac41a9}
  {\bibfield  {journal} {\bibinfo  {journal} {Journal of Physics
  Communications}\ }\textbf {\bibinfo {volume} {5}},\ \bibinfo {pages} {125012}
  (\bibinfo {year} {2021})}\BibitemShut {NoStop}%
\bibitem [{\citenamefont {Richards}\ \emph {et~al.}(1997)\citenamefont
  {Richards}, \citenamefont {Wolf},\ and\ \citenamefont
  {Gabor}}]{Richards1959EM-diffraction}%
  \BibitemOpen
  \bibfield  {author} {\bibinfo {author} {\bibfnamefont {B.}~\bibnamefont
  {Richards}}, \bibinfo {author} {\bibfnamefont {E.}~\bibnamefont {Wolf}},\
  and\ \bibinfo {author} {\bibfnamefont {D.}~\bibnamefont {Gabor}},\ }\bibfield
   {title} {\bibinfo {title} {Electromagnetic diffraction in optical systems,
  ii. structure of the image field in an aplanatic system},\ }\href
  {https://doi.org/10.1098/rspa.1959.0200} {\bibfield  {journal} {\bibinfo
  {journal} {Proceedings of the Royal Society of London. Series A. Mathematical
  and Physical Sciences}\ }\textbf {\bibinfo {volume} {253}},\ \bibinfo {pages}
  {358} (\bibinfo {year} {1997})}\BibitemShut {NoStop}%
\bibitem [{\citenamefont {Bekshaev}\ \emph {et~al.}(2011)\citenamefont
  {Bekshaev}, \citenamefont {Bliokh},\ and\ \citenamefont
  {Soskin}}]{Bekshaev2011Internal}%
  \BibitemOpen
  \bibfield  {author} {\bibinfo {author} {\bibfnamefont {A.}~\bibnamefont
  {Bekshaev}}, \bibinfo {author} {\bibfnamefont {K.~Y.}\ \bibnamefont
  {Bliokh}},\ and\ \bibinfo {author} {\bibfnamefont {M.}~\bibnamefont
  {Soskin}},\ }\bibfield  {title} {\bibinfo {title} {Internal flows and energy
  circulation in light beams},\ }\href
  {https://doi.org/10.1088/2040-8978/13/5/053001} {\bibfield  {journal}
  {\bibinfo  {journal} {Journal of Optics}\ }\textbf {\bibinfo {volume} {13}},\
  \bibinfo {pages} {053001} (\bibinfo {year} {2011})}\BibitemShut {NoStop}%
\bibitem [{\citenamefont {Fernandez-Corbaton}\ \emph
  {et~al.}(2013)\citenamefont {Fernandez-Corbaton}, \citenamefont
  {Zambrana-Puyalto}, \citenamefont {Tischler}, \citenamefont {Vidal},
  \citenamefont {Juan},\ and\ \citenamefont
  {Molina-Terriza}}]{Fernandez2013EM_duality}%
  \BibitemOpen
  \bibfield  {author} {\bibinfo {author} {\bibfnamefont {I.}~\bibnamefont
  {Fernandez-Corbaton}}, \bibinfo {author} {\bibfnamefont {X.}~\bibnamefont
  {Zambrana-Puyalto}}, \bibinfo {author} {\bibfnamefont {N.}~\bibnamefont
  {Tischler}}, \bibinfo {author} {\bibfnamefont {X.}~\bibnamefont {Vidal}},
  \bibinfo {author} {\bibfnamefont {M.~L.}\ \bibnamefont {Juan}},\ and\
  \bibinfo {author} {\bibfnamefont {G.}~\bibnamefont {Molina-Terriza}},\
  }\bibfield  {title} {\bibinfo {title} {Electromagnetic duality symmetry and
  helicity conservation for the macroscopic maxwell's equations},\ }\href
  {https://doi.org/10.1103/PhysRevLett.111.060401} {\bibfield  {journal}
  {\bibinfo  {journal} {Phys Rev Lett}\ }\textbf {\bibinfo {volume} {111}},\
  \bibinfo {pages} {060401} (\bibinfo {year} {2013})}\BibitemShut {NoStop}%
\bibitem [{SM()}]{SM}%
  \BibitemOpen
  \bibfield  {title} {\bibinfo {title} {See {Supplemental Material} [url] for
  the detailed derivation of the decomposition for transverse spins, and the
  analytic results for several structured vector fields including cosine waves,
  plane waves, elliptic polarized waves, two-wave interference, {B}essel beams
  and {A}iry beams. also the acoustic bessel beam is also presented in sec.
  {VIII}.},\ }\href@noop {} {\ }\BibitemShut {NoStop}%
\bibitem [{Note1()}]{Note1}%
  \BibitemOpen
  \bibinfo {note} {Note that our $\protect \textbf {s}_{\protect \rm p}$ is
  defined twice of that from ~\cite {Vernon2024}.}\BibitemShut {Stop}%
\bibitem [{\citenamefont {McGloin}\ and\ \citenamefont
  {Dholakia}(2005)}]{Mcgloin2005}%
  \BibitemOpen
  \bibfield  {author} {\bibinfo {author} {\bibfnamefont {D.}~\bibnamefont
  {McGloin}}\ and\ \bibinfo {author} {\bibfnamefont {K.}~\bibnamefont
  {Dholakia}},\ }\bibfield  {title} {\bibinfo {title} {Bessel beams:
  diffraction in a new light},\ }\href
  {https://doi.org/10.1080/0010751042000275259} {\bibfield  {journal} {\bibinfo
   {journal} {Contemporary Physics}\ }\textbf {\bibinfo {volume} {46}},\
  \bibinfo {pages} {15} (\bibinfo {year} {2005})}\BibitemShut {NoStop}%
\bibitem [{\citenamefont {Mishra}(1991)}]{Mishra1991}%
  \BibitemOpen
  \bibfield  {author} {\bibinfo {author} {\bibfnamefont {S.}~\bibnamefont
  {Mishra}},\ }\bibfield  {title} {\bibinfo {title} {A vector wave analysis of
  a bessel beam},\ }\href {https://doi.org/10.1016/0030-4018(91)90386-R}
  {\bibfield  {journal} {\bibinfo  {journal} {Optics Communications}\ }\textbf
  {\bibinfo {volume} {85}},\ \bibinfo {pages} {159} (\bibinfo {year}
  {1991})}\BibitemShut {NoStop}%
\bibitem [{\citenamefont {Bouchal}\ and\ \citenamefont
  {Oliv{\'\i}k}(1995)}]{Bouchal1995}%
  \BibitemOpen
  \bibfield  {author} {\bibinfo {author} {\bibfnamefont {Z.}~\bibnamefont
  {Bouchal}}\ and\ \bibinfo {author} {\bibfnamefont {M.}~\bibnamefont
  {Oliv{\'\i}k}},\ }\bibfield  {title} {\bibinfo {title} {Non-diffractive
  vector bessel beams},\ }\href {https://doi.org/10.1080/09500349514551361}
  {\bibfield  {journal} {\bibinfo  {journal} {Journal of Modern Optics}\
  }\textbf {\bibinfo {volume} {42}},\ \bibinfo {pages} {1555} (\bibinfo {year}
  {1995})}\BibitemShut {NoStop}%
\bibitem [{\citenamefont {Zhang}\ \emph {et~al.}(2024)\citenamefont {Zhang},
  \citenamefont {Liu}, \citenamefont {Hu}, \citenamefont {Lin}, \citenamefont
  {Zeng}, \citenamefont {Zhang}, \citenamefont {Li}, \citenamefont {Chen},\
  and\ \citenamefont {Fu}}]{ZhangX2024Ph_spin-orbit}%
  \BibitemOpen
  \bibfield  {author} {\bibinfo {author} {\bibfnamefont {X.}~\bibnamefont
  {Zhang}}, \bibinfo {author} {\bibfnamefont {G.}~\bibnamefont {Liu}}, \bibinfo
  {author} {\bibfnamefont {Y.}~\bibnamefont {Hu}}, \bibinfo {author}
  {\bibfnamefont {H.}~\bibnamefont {Lin}}, \bibinfo {author} {\bibfnamefont
  {Z.}~\bibnamefont {Zeng}}, \bibinfo {author} {\bibfnamefont {X.}~\bibnamefont
  {Zhang}}, \bibinfo {author} {\bibfnamefont {Z.}~\bibnamefont {Li}}, \bibinfo
  {author} {\bibfnamefont {Z.}~\bibnamefont {Chen}},\ and\ \bibinfo {author}
  {\bibfnamefont {S.}~\bibnamefont {Fu}},\ }\bibfield  {title} {\bibinfo
  {title} {Photonic spin-orbit coupling induced by deep-subwavelength
  structured light},\ }\href {https://doi.org/10.1103/PhysRevA.109.023522}
  {\bibfield  {journal} {\bibinfo  {journal} {Physical Review A}\ }\textbf
  {\bibinfo {volume} {109}},\ \bibinfo {pages} {023522} (\bibinfo {year}
  {2024})}\BibitemShut {NoStop}%
\bibitem [{\citenamefont {Du}\ \emph {et~al.}(2019)\citenamefont {Du},
  \citenamefont {Yang}, \citenamefont {Zayats},\ and\ \citenamefont
  {Yuan}}]{DuL2019Deep-sub}%
  \BibitemOpen
  \bibfield  {author} {\bibinfo {author} {\bibfnamefont {L.}~\bibnamefont
  {Du}}, \bibinfo {author} {\bibfnamefont {A.}~\bibnamefont {Yang}}, \bibinfo
  {author} {\bibfnamefont {A.~V.}\ \bibnamefont {Zayats}},\ and\ \bibinfo
  {author} {\bibfnamefont {X.}~\bibnamefont {Yuan}},\ }\bibfield  {title}
  {\bibinfo {title} {Deep-subwavelength features of photonic skyrmions in a
  confined electromagnetic field with orbital angular momentum},\ }\href
  {https://www.nature.com/articles/s41567-019-0487-7} {\bibfield  {journal}
  {\bibinfo  {journal} {Nature Physics}\ }\textbf {\bibinfo {volume} {15}},\
  \bibinfo {pages} {650} (\bibinfo {year} {2019})}\BibitemShut {NoStop}%
\bibitem [{\citenamefont {Agrawal}\ and\ \citenamefont
  {Pattanayak}(1979)}]{Agrawal1979}%
  \BibitemOpen
  \bibfield  {author} {\bibinfo {author} {\bibfnamefont {G.~P.}\ \bibnamefont
  {Agrawal}}\ and\ \bibinfo {author} {\bibfnamefont {D.~N.}\ \bibnamefont
  {Pattanayak}},\ }\bibfield  {title} {\bibinfo {title} {Gaussian beam
  propagation beyond the paraxial approximation},\ }\href
  {https://doi.org/10.1364/JOSA.69.000575} {\bibfield  {journal} {\bibinfo
  {journal} {Journal of the Optical Society of America}\ }\textbf {\bibinfo
  {volume} {69}},\ \bibinfo {pages} {575} (\bibinfo {year} {1979})}\BibitemShut
  {NoStop}%
\bibitem [{\citenamefont {Alda}(2003)}]{Alda2003}%
  \BibitemOpen
  \bibfield  {author} {\bibinfo {author} {\bibfnamefont {J.}~\bibnamefont
  {Alda}},\ }\bibfield  {title} {\bibinfo {title} {Laser and gaussian beam
  propagation and transformation},\ }\href
  {https://sites.unimi.it/aqm/wp-content/uploads/JAlda-2003.pdf} {\bibfield
  {journal} {\bibinfo  {journal} {Encyclopedia of optical engineering}\ }
  (\bibinfo {year} {2003})}\BibitemShut {NoStop}%
\bibitem [{\citenamefont {Siviloglou}\ and\ \citenamefont
  {Christodoulides}(2007)}]{Siviloglou2007}%
  \BibitemOpen
  \bibfield  {author} {\bibinfo {author} {\bibfnamefont {G.~A.}\ \bibnamefont
  {Siviloglou}}\ and\ \bibinfo {author} {\bibfnamefont {D.~N.}\ \bibnamefont
  {Christodoulides}},\ }\bibfield  {title} {\bibinfo {title} {Accelerating
  finite energy airy beams},\ }\href {https://doi.org/10.1364/OL.32.000979}
  {\bibfield  {journal} {\bibinfo  {journal} {Optics Letters}\ }\textbf
  {\bibinfo {volume} {32}},\ \bibinfo {pages} {979} (\bibinfo {year}
  {2007})}\BibitemShut {NoStop}%
\bibitem [{\citenamefont {Siviloglou}\ \emph {et~al.}(2007)\citenamefont
  {Siviloglou}, \citenamefont {Broky}, \citenamefont {Dogariu},\ and\
  \citenamefont {Christodoulides}}]{Siviloglou2007observation}%
  \BibitemOpen
  \bibfield  {author} {\bibinfo {author} {\bibfnamefont {G.}~\bibnamefont
  {Siviloglou}}, \bibinfo {author} {\bibfnamefont {J.}~\bibnamefont {Broky}},
  \bibinfo {author} {\bibfnamefont {A.}~\bibnamefont {Dogariu}},\ and\ \bibinfo
  {author} {\bibfnamefont {D.}~\bibnamefont {Christodoulides}},\ }\bibfield
  {title} {\bibinfo {title} {Observation of accelerating airy beams},\ }\href
  {https://journals.aps.org/prl/abstract/10.1103/PhysRevLett.99.213901}
  {\bibfield  {journal} {\bibinfo  {journal} {Physical Review Letters}\
  }\textbf {\bibinfo {volume} {99}},\ \bibinfo {pages} {213901} (\bibinfo
  {year} {2007})}\BibitemShut {NoStop}%
\bibitem [{\citenamefont {Yin}\ \emph {et~al.}(2020)\citenamefont {Yin},
  \citenamefont {Shi}, \citenamefont {Du},\ and\ \citenamefont
  {Yuan}}]{YinX2020Spin-resolved}%
  \BibitemOpen
  \bibfield  {author} {\bibinfo {author} {\bibfnamefont {X.}~\bibnamefont
  {Yin}}, \bibinfo {author} {\bibfnamefont {P.}~\bibnamefont {Shi}}, \bibinfo
  {author} {\bibfnamefont {L.}~\bibnamefont {Du}},\ and\ \bibinfo {author}
  {\bibfnamefont {X.}~\bibnamefont {Yuan}},\ }\bibfield  {title} {\bibinfo
  {title} {Spin-resolved near-field scanning optical microscopy for mapping of
  the spin angular momentum distribution of focused beams},\ }\bibfield
  {journal} {\bibinfo  {journal} {Applied Physics Letters}\ }\textbf {\bibinfo
  {volume} {116}},\ \href {https://doi.org/10.1063/5.0004750}
  {10.1063/5.0004750} (\bibinfo {year} {2020})\BibitemShut {NoStop}%
\bibitem [{\citenamefont {Taneja}\ \emph {et~al.}(2021)\citenamefont {Taneja},
  \citenamefont {Paul},\ and\ \citenamefont {Pavan~Kumar}}]{Taneja2021t_spin}%
  \BibitemOpen
  \bibfield  {author} {\bibinfo {author} {\bibfnamefont {C.}~\bibnamefont
  {Taneja}}, \bibinfo {author} {\bibfnamefont {D.}~\bibnamefont {Paul}},\ and\
  \bibinfo {author} {\bibfnamefont {G.~V.}\ \bibnamefont {Pavan~Kumar}},\
  }\bibfield  {title} {\bibinfo {title} {Experimental observation of transverse
  spin of plasmon polaritons in a single crystalline silver nanowire},\ }\href
  {https://doi.org/10.1063/5.0055788} {\bibfield  {journal} {\bibinfo
  {journal} {Applied Physics Letters}\ }\textbf {\bibinfo {volume} {119}},\
  \bibinfo {pages} {161108} (\bibinfo {year} {2021})}\BibitemShut {NoStop}%
\bibitem [{\citenamefont {Frischwasser}\ \emph {et~al.}(2021)\citenamefont
  {Frischwasser}, \citenamefont {Cohen}, \citenamefont {Kher-Alden},
  \citenamefont {Dolev}, \citenamefont {Tsesses},\ and\ \citenamefont
  {Bartal}}]{Frischwasse2021Real-time}%
  \BibitemOpen
  \bibfield  {author} {\bibinfo {author} {\bibfnamefont {K.}~\bibnamefont
  {Frischwasser}}, \bibinfo {author} {\bibfnamefont {K.}~\bibnamefont {Cohen}},
  \bibinfo {author} {\bibfnamefont {J.}~\bibnamefont {Kher-Alden}}, \bibinfo
  {author} {\bibfnamefont {S.}~\bibnamefont {Dolev}}, \bibinfo {author}
  {\bibfnamefont {S.}~\bibnamefont {Tsesses}},\ and\ \bibinfo {author}
  {\bibfnamefont {G.}~\bibnamefont {Bartal}},\ }\bibfield  {title} {\bibinfo
  {title} {Real-time sub-wavelength imaging of surface waves with nonlinear
  near-field optical microscopy},\ }\href
  {https://doi.org/10.1038/s41566-021-00782-2} {\bibfield  {journal} {\bibinfo
  {journal} {Nature Photonics}\ }\textbf {\bibinfo {volume} {15}},\ \bibinfo
  {pages} {442} (\bibinfo {year} {2021})}\BibitemShut {NoStop}%
\bibitem [{\citenamefont {Shi}\ \emph {et~al.}(2022{\natexlab{b}})\citenamefont
  {Shi}, \citenamefont {Zhu}, \citenamefont {Liu}, \citenamefont {Tsai},
  \citenamefont {Zhang}, \citenamefont {Wang}, \citenamefont {Chan},
  \citenamefont {Wu}, \citenamefont {Zayats}, \citenamefont {Nori} \emph
  {et~al.}}]{ShiY2022stable}%
  \BibitemOpen
  \bibfield  {author} {\bibinfo {author} {\bibfnamefont {Y.}~\bibnamefont
  {Shi}}, \bibinfo {author} {\bibfnamefont {T.}~\bibnamefont {Zhu}}, \bibinfo
  {author} {\bibfnamefont {J.}~\bibnamefont {Liu}}, \bibinfo {author}
  {\bibfnamefont {D.}~\bibnamefont {Tsai}}, \bibinfo {author} {\bibfnamefont
  {H.}~\bibnamefont {Zhang}}, \bibinfo {author} {\bibfnamefont
  {S.}~\bibnamefont {Wang}}, \bibinfo {author} {\bibfnamefont {C.~T.}\
  \bibnamefont {Chan}}, \bibinfo {author} {\bibfnamefont {P.}~\bibnamefont
  {Wu}}, \bibinfo {author} {\bibfnamefont {A.}~\bibnamefont {Zayats}}, \bibinfo
  {author} {\bibfnamefont {F.}~\bibnamefont {Nori}}, \emph {et~al.},\
  }\bibfield  {title} {\bibinfo {title} {Stable optical lateral forces from
  inhomogeneities of the spin angular momentum},\ }\href
  {https://www.science.org/doi/full/10.1126/sciadv.abn2291} {\bibfield
  {journal} {\bibinfo  {journal} {Science {A}dvances}\ }\textbf {\bibinfo
  {volume} {8}},\ \bibinfo {pages} {eabn2291} (\bibinfo {year}
  {2022}{\natexlab{b}})}\BibitemShut {NoStop}%
\bibitem [{\citenamefont {Peng}\ \emph {et~al.}(2022)\citenamefont {Peng},
  \citenamefont {Ren}, \citenamefont {Liu}, \citenamefont {Lan}, \citenamefont
  {Xu}, \citenamefont {Ye}, \citenamefont {Sun}, \citenamefont {Xu},
  \citenamefont {Chen},\ and\ \citenamefont {Zhang}}]{PengL2022spin_Hall}%
  \BibitemOpen
  \bibfield  {author} {\bibinfo {author} {\bibfnamefont {L.}~\bibnamefont
  {Peng}}, \bibinfo {author} {\bibfnamefont {H.}~\bibnamefont {Ren}}, \bibinfo
  {author} {\bibfnamefont {Y.-C.}\ \bibnamefont {Liu}}, \bibinfo {author}
  {\bibfnamefont {T.}~\bibnamefont {Lan}}, \bibinfo {author} {\bibfnamefont
  {K.}~\bibnamefont {Xu}}, \bibinfo {author} {\bibfnamefont {D.}~\bibnamefont
  {Ye}}, \bibinfo {author} {\bibfnamefont {H.}~\bibnamefont {Sun}}, \bibinfo
  {author} {\bibfnamefont {S.}~\bibnamefont {Xu}}, \bibinfo {author}
  {\bibfnamefont {H.-S.}\ \bibnamefont {Chen}},\ and\ \bibinfo {author}
  {\bibfnamefont {S.}~\bibnamefont {Zhang}},\ }\bibfield  {title} {\bibinfo
  {title} {Spin {Hall} effect of transversely spinning light},\ }\href
  {https://doi.org/10.1126/sciadv.abo6033} {\bibfield  {journal} {\bibinfo
  {journal} {Science {A}dvances}\ }\textbf {\bibinfo {volume} {8}},\ \bibinfo
  {pages} {eabo6033} (\bibinfo {year} {2022})}\BibitemShut {NoStop}%
\bibitem [{\citenamefont {Dai}\ \emph {et~al.}(2022)\citenamefont {Dai},
  \citenamefont {Zhou}, \citenamefont {Ghosh}, \citenamefont {Kapoor},
  \citenamefont {D{\k{a}}browski}, \citenamefont {Kubo}, \citenamefont
  {Huang},\ and\ \citenamefont {Petek}}]{DaiY2022Ultrafast}%
  \BibitemOpen
  \bibfield  {author} {\bibinfo {author} {\bibfnamefont {Y.}~\bibnamefont
  {Dai}}, \bibinfo {author} {\bibfnamefont {Z.}~\bibnamefont {Zhou}}, \bibinfo
  {author} {\bibfnamefont {A.}~\bibnamefont {Ghosh}}, \bibinfo {author}
  {\bibfnamefont {K.}~\bibnamefont {Kapoor}}, \bibinfo {author} {\bibfnamefont
  {M.}~\bibnamefont {D{\k{a}}browski}}, \bibinfo {author} {\bibfnamefont
  {A.}~\bibnamefont {Kubo}}, \bibinfo {author} {\bibfnamefont {C.-B.}\
  \bibnamefont {Huang}},\ and\ \bibinfo {author} {\bibfnamefont
  {H.}~\bibnamefont {Petek}},\ }\bibfield  {title} {\bibinfo {title} {Ultrafast
  microscopy of a twisted plasmonic spin skyrmion},\ }\bibfield  {journal}
  {\bibinfo  {journal} {Applied Physics Reviews}\ }\textbf {\bibinfo {volume}
  {9}},\ \href {https://doi.org/10.1063/5.0084482} {10.1063/5.0084482}
  (\bibinfo {year} {2022})\BibitemShut {NoStop}%
\bibitem [{\citenamefont {Chen}\ \emph {et~al.}(2022)\citenamefont {Chen},
  \citenamefont {Zhang}, \citenamefont {Liu}, \citenamefont {Meng},
  \citenamefont {Dudley},\ and\ \citenamefont {Lu}}]{ChenW2022Time}%
  \BibitemOpen
  \bibfield  {author} {\bibinfo {author} {\bibfnamefont {W.}~\bibnamefont
  {Chen}}, \bibinfo {author} {\bibfnamefont {W.}~\bibnamefont {Zhang}},
  \bibinfo {author} {\bibfnamefont {Y.}~\bibnamefont {Liu}}, \bibinfo {author}
  {\bibfnamefont {F.~C.}\ \bibnamefont {Meng}}, \bibinfo {author}
  {\bibfnamefont {J.~M.}\ \bibnamefont {Dudley}},\ and\ \bibinfo {author}
  {\bibfnamefont {Y.~Q.}\ \bibnamefont {Lu}},\ }\bibfield  {title} {\bibinfo
  {title} {Time diffraction-free transverse orbital angular momentum beams},\
  }\href {https://doi.org/10.1038/s41467-022-31623-7} {\bibfield  {journal}
  {\bibinfo  {journal} {Nat Commun}\ }\textbf {\bibinfo {volume} {13}},\
  \bibinfo {pages} {4021} (\bibinfo {year} {2022})}\BibitemShut {NoStop}%
\bibitem [{\citenamefont {Bliokh}\ and\ \citenamefont
  {Nori}(2019{\natexlab{a}})}]{Bliokh2019Spin}%
  \BibitemOpen
  \bibfield  {author} {\bibinfo {author} {\bibfnamefont {K.~Y.}\ \bibnamefont
  {Bliokh}}\ and\ \bibinfo {author} {\bibfnamefont {F.}~\bibnamefont {Nori}},\
  }\bibfield  {title} {\bibinfo {title} {Spin and orbital angular momenta of
  acoustic beams},\ }\bibfield  {journal} {\bibinfo  {journal} {Physical Review
  B}\ }\textbf {\bibinfo {volume} {99}},\ \href
  {https://doi.org/10.1103/PhysRevB.99.174310} {10.1103/PhysRevB.99.174310}
  (\bibinfo {year} {2019}{\natexlab{a}})\BibitemShut {NoStop}%
\bibitem [{\citenamefont {Bliokh}\ and\ \citenamefont
  {Nori}(2022)}]{Bliokh2022Erratum}%
  \BibitemOpen
  \bibfield  {author} {\bibinfo {author} {\bibfnamefont {K.~Y.}\ \bibnamefont
  {Bliokh}}\ and\ \bibinfo {author} {\bibfnamefont {F.}~\bibnamefont {Nori}},\
  }\bibfield  {title} {\bibinfo {title} {Erratum: {S}pin and orbital angular
  momenta of acoustic beams [{Phys. Rev. B} 99, 174310 (2019)]},\ }\bibfield
  {journal} {\bibinfo  {journal} {Physical Review B}\ }\textbf {\bibinfo
  {volume} {105}},\ \href {https://doi.org/10.1103/PhysRevB.105.219901}
  {10.1103/PhysRevB.105.219901} (\bibinfo {year} {2022})\BibitemShut {NoStop}%
\bibitem [{\citenamefont {Bliokh}\ and\ \citenamefont
  {Nori}(2019{\natexlab{b}})}]{Bliokh2019t-spin}%
  \BibitemOpen
  \bibfield  {author} {\bibinfo {author} {\bibfnamefont {K.~Y.}\ \bibnamefont
  {Bliokh}}\ and\ \bibinfo {author} {\bibfnamefont {F.}~\bibnamefont {Nori}},\
  }\bibfield  {title} {\bibinfo {title} {Transverse spin and surface waves in
  acoustic metamaterials},\ }\href {https://doi.org/10.1103/PhysRevB.99.020301}
  {\bibfield  {journal} {\bibinfo  {journal} {Phys. Rev. B}\ }\textbf {\bibinfo
  {volume} {99}},\ \bibinfo {pages} {020301} (\bibinfo {year}
  {2019}{\natexlab{b}})}\BibitemShut {NoStop}%
\bibitem [{\citenamefont {Bliokh}\ \emph {et~al.}(2022)\citenamefont {Bliokh},
  \citenamefont {Punzmann}, \citenamefont {Xia}, \citenamefont {Nori},\ and\
  \citenamefont {Shats}}]{Bliokh2022Field}%
  \BibitemOpen
  \bibfield  {author} {\bibinfo {author} {\bibfnamefont {K.~Y.}\ \bibnamefont
  {Bliokh}}, \bibinfo {author} {\bibfnamefont {H.}~\bibnamefont {Punzmann}},
  \bibinfo {author} {\bibfnamefont {H.}~\bibnamefont {Xia}}, \bibinfo {author}
  {\bibfnamefont {F.}~\bibnamefont {Nori}},\ and\ \bibinfo {author}
  {\bibfnamefont {M.}~\bibnamefont {Shats}},\ }\bibfield  {title} {\bibinfo
  {title} {Field theory spin and momentum in water waves},\ }\href
  {https://doi.org/10.1126/sciadv.abm1295} {\bibfield  {journal} {\bibinfo
  {journal} {Science Advances}\ }\textbf {\bibinfo {volume} {8}},\ \bibinfo
  {pages} {eabm1295} (\bibinfo {year} {2022})}\BibitemShut {NoStop}%
\bibitem [{\citenamefont {Long}\ \emph {et~al.}(2023)\citenamefont {Long},
  \citenamefont {Yang}, \citenamefont {Chen},\ and\ \citenamefont
  {Ren}}]{LongY2023}%
  \BibitemOpen
  \bibfield  {author} {\bibinfo {author} {\bibfnamefont {Y.}~\bibnamefont
  {Long}}, \bibinfo {author} {\bibfnamefont {C.}~\bibnamefont {Yang}}, \bibinfo
  {author} {\bibfnamefont {H.}~\bibnamefont {Chen}},\ and\ \bibinfo {author}
  {\bibfnamefont {J.}~\bibnamefont {Ren}},\ }\bibfield  {title} {\bibinfo
  {title} {Universal geometric relations of acoustic spin, energy flux, and
  reactive power},\ }\bibfield  {journal} {\bibinfo  {journal} {Physical Review
  Applied}\ }\textbf {\bibinfo {volume} {19}},\ \href
  {https://doi.org/10.1103/PhysRevApplied.19.064053}
  {10.1103/PhysRevApplied.19.064053} (\bibinfo {year} {2023})\BibitemShut
  {NoStop}%
\end{thebibliography}%
\end{document}